\documentclass[10pt]{revtex4}
\usepackage{amssymb}
\usepackage{latexsym}
\usepackage{epsfig}
\usepackage{amsmath}
\begin{document}
\title{Statefinder diagnosis for holographic dark energy in the DGP braneworld}
\author{S. Ghaffari$^{1}$, A. Sheykhi$^{1,2}$
\thanks{asheykhi@shirazu.ac.ir} and M. H. Dehghani, $^{1,2}$\thanks{mhd@shirazu.ac.ir}
}
\address{$^1$  Physics Department and Biruni Observatory, College of
Sciences, Shiraz University, Shiraz 71454, Iran\\
         $^2$  Research Institute for Astronomy and Astrophysics of Maragha
         (RIAAM), P.O. Box 55134-441, Maragha, Iran}
\begin{abstract}
Many dark energy (DE) models have been proposed, in recent years,
to explain acceleration of the Universe expansion. It seems
necessary to discriminate the various DE models in order to check
the viability of each model. Statefinder diagnostic is a useful
method which can differentiate various DE models. In this paper,
we investigate the statefinder diagnosis parameters for the
holographic dark energy (HDE) model in two cosmological setup.
First, we study statefinder diagnosis for HDE in the context of
flat Friedmann-Robertson-Walker (FRW) Universe in Einstein
gravity. Then, we extend our study to the DGP braneworld
framework. As system's IR cutoff we chose the Hubble radius and
the Granda-Oliveros cutoff inspired by Ricci scalar curvature. We
plot the evolution of statefinder parameres $\{r,s\}$ in terms of
the redshift parameter $z$. We also compare the results with those
obtained for statefinder diagnosis parameters of other DE models,
in particular with $\Lambda$CDM model.
\end{abstract}
\maketitle
\section{Introduction}

Recent astronomical observations such as type Ia supernovae (SNIa)
\cite{Riess}, large scale structure (LSS) \cite{COL2001}, and the
cosmic microwave background (CMB) anisotropies \cite{HAN2000}
confirm that our Universe is currently undergoing a phase of
accelerated expansion. DE which is responsible for this
acceleration has anti-gravity nature and hence push the Universe
to accelerate. Although the nature of such a DE is still unknown,
many candidates have been proposed for DE in the literatures. The
first and simplest candidate for DE is the cosmological constant
with equation of state (EoS) parameter $\omega=-1$. But
unfortunately, it suffers from some serious problems such as
fine-tuning and coincidence problems. A comprehensive but
incomplete list of DE scenarios include scalar fields (such as
quintessence \cite{Wetterich}, phantom (ghost) field
\cite{Caldwell}, K-essence \cite{Chiba}), the DE models
including Chaplygin gas \cite{Kamenshchik}, agegraphic DE models
\cite{Cai1,Shey2}, HDE model \cite{Cohen1,Hsu,Li,pav1,Shey1} and
so on.

The HDE model, which has got a lot of attentions in recent years,
is based on Cohen et al. \cite{Cohen1} works who discussed that in
quantum field theory a short distance cutoff could be related to a
long distance cutoff (IR cutoff) due to the limit set by black
hole formation. If the quantum zero-point energy density is due to
a short distance cutoff, then the total energy in a region of size
$L$ should not exceed the mass of a black hole of the same size,
namely $L^3\rho_D\leq L M_p^2$. The largest $L$ is the one
saturating this inequality and therefore we obtain the HDE density as
\cite{Cohen1}
\begin{equation}\label{HDE}
\rho_D=3c^2M_p^2L^{-2},
\end{equation}
where $c$ is a numerical constant and $M^2_p = 8\pi G$ is the
reduced Planck mass. In the literatures, various IR cutoffs have
been considered, which lead to various models of HDE
\cite{Horava,Fischler,Nojiri,Gao}. Recently, Granda and Oliveros
(GO) have proposed a new IR cutoff which includes time derivative
of the Hubble parameter and is the formal generalization of the
Ricci scalar curvature, $L=(\alpha \dot{H}+\beta H^2)^{-1/2}$
\cite{Granda}. The advantages of GO cutoff is that the presence of
the event horizon is not presumed in this model, so that the
causality problem can be avoided \cite{Granda}. Besides, the fine
tuning problem can be solved in this model \cite{Granda}.

On the other hand, the expansion rate of the Universe is explained
by the Hubble parameter $H=\dot{a}/a$, where $a=a(t)$ is the scale
factor of the Universe, while the rate of the acceleration or
deceleration of the Universe expansion is described by the
deceleration parameter,
\begin{equation}\label{q}
q=-\frac{a\ddot{a}}{\dot{a}^2}=-\frac{\ddot{a}}{aH^2}.
\end{equation}
However, the Hubble parameter $H$ and the deceleration parameter
$q$  cannot discriminate various DE models since all DE models
leads to $H>0$ and $\ddot{a}>0$ or $q<0$. In addition, the
remarkable increase in the accuracy of cosmological observational
data during the last few years, compel us to advance beyond these
two important quantities. A question then arise: how  can we
discriminate and classify various models of DE? In order to answer
this question, Sahni, et. al., \cite{Sahni} and  Alam, et al.,
\cite{Alam}, proposed a new geometrical diagnostic pair for DE.
This diagnostic is constructed from scale factor $a(t)$ and its
derivatives up to the third order. The statefinder pair $\{r,s\}$
is defined as \cite{Sahni}
\begin{equation}\label{r}
r=\frac{\dddot{a}}{aH^3},~~~~~~~~~~~~~~ s=\frac{r-1}{3(q-1/2)}.
\end{equation}

The statefinder pair is a geometrical diagnostic in the sense that
it is constructed from a spacetime metric directly, and it is more
universal than physical variables which depend on the properties of
physical fields describing DE, because physical variables are, of
course, model-dependent. Usually one can plot the trajectories in
the $r-s$ plane corresponding to different DE models to see the
qualitatively different behaviors of them. For flat $\Lambda$CDM
scenario the statefinder pair is $\{r,s\}=\{1,0\}$ \cite{Feng}.
This fixed point provides a good way of establishing the
"distance" of any given DE model from $\Lambda$CDM model. It was
shown that the statefinder pair $\{r,s\}$ can also be related to
the EoS parameter of DE and its first time derivative
\cite{Sahni}. The statefinder pair is calculated for a number of
existing models of DE having both constant and variable EoS
parameter. It was argued that the statefinder can successfully
differentiate between a wide variety of DE models including the
cosmological constant, quintessence, the Chaplygin gas and
interacting DE models (see \cite{state, state1} and references
therein). In this paper we would like to study the statefinder
pair parameters $\{r,s\}$ for HDE in standard cosmology as well as
DGP braneworld scenario.

This paper is outlined as follows. In the next section we
investigate statefinder for HDE in standard cosmology with
considering Hubble radius and GO cutoff as systems's IR cutoffs.
We also plot the related figures which show the evolution of
statefinder in each case. In section III, we extended our study to
DGP braneworld and calculate statefinder pair $\{r,s\}$ for HDE in
a flat FRW Universe on the brane. The last section is devoted to
conclusions and discussions.
%%%%%%%%%%%%%%%%%%%%%%%%%%%%%%%%%%%%%%%%%%%%%%%%%%%%%%%%%%%%%%%%%%%%%%%%%%
\section{HDE in standard cosmology}
Consider a homogeneous and isotropic FRW which has the following
metric
\begin{equation}
{\rm d}s^2=-{\rm d}t^2+a^2(t)\left(\frac{{\rm d}r^2}{1-kr^2}+r^2{\rm
d}\Omega^2\right),\label{metric}
\end{equation}
where $k=0,1,-1$ represent a flat, closed and open FRW Universe,
respectively. Due to the fact that the density of baryonic matter
is much smaller than the energy densities of DM and DE, we
ignore the baryonic matter throughout the paper. Thus, Friedmann equation in spatially flat FRW
Universe may be written as
\begin{equation}\label{Friedeq1}
H^2=\frac{1}{3M_p^2}(\rho_m+\rho_D),
\end{equation}
where $\rho_m$ and $\rho_D$ are the energy densities of DM and DE,
respectively. From equation (\ref{Friedeq1}) we can write
\begin{equation}\label{Friedeq2}
\Omega_m+\Omega_D=1.
\end{equation}
where we have used the following dimensionless energy densities
definitions,
\begin{equation}\label{Friedeq3}
\Omega_m=\frac{\rho_m}{3M_p^2H^2},~~~~~~~~~~~~~~~~\Omega_D=\frac{\rho_D}{3M_p^2H^2}.
\end{equation}
Recent observational evidences provided by the galaxy clusters
supports the interaction between DE and DM \cite{Bertolami}. In
this case, the energy densities of DE and DM no longer satisfy
independent conservation laws. They obey instead
\begin{eqnarray}\label{ConserveCDMQ}
&&\dot{\rho}_m+3H\rho_m=Q,\\
&&\dot{\rho}_D+3H(1+\omega_D)\rho_D=-Q,\label{ConserveDEQ}
\end{eqnarray}
where $\omega_D=p_D/\rho_D$ is the EoS parameter of HDE and $Q$
describes the interaction between DE and DM. The choice of the
interaction between both components was to get a scaling solution
to the coincidence problem such that the Universe approaches a
stationary stage in which the ratio of DE and DM becomes a
constant \cite{Hu}. Here, we choose $Q = 3b^2H(\rho_D+\rho_m)$ as
an interaction term with $b^2$ being a coupling constant. This
expression for the interaction term was first introduced in the
study of the suitable coupling between a quintessence scalar field
and a pressureless cold DM field \cite{Amendola,Zimdahl}.

First, we choose  Hubble radius as IR cutoff, $L=H^{-1}$, thus the
the hologeraphic energy density reads
\begin{equation}\label{HubbleHDE}
\rho_D=3c^2M_p^2H^2.
\end{equation}
If we consider $c$ as a constant, then from definition
(\ref{Friedeq3}) the dimensionless DE density becomes a constant,
namely $\Omega_D=c^2$. In this case, without interaction between
DE and DM the accelerated expansion of the Universe cannot be
achieved and we get a wrong EoS, namely that for dust,
$\omega_D=0$ \cite{Li}. However, as soon as the interaction
between two dark component is taken into account, the accelerated
expansion can be achieved and implies a constant ratio of the
energy densities of both components, thus solving the coincidence
problem \cite{pav1}.

Taking the time derivative of the Friedmann equation
(\ref{Friedeq1}) and using Eqs.
(\ref{Friedeq2})-(\ref{ConserveDEQ}), we get
\begin{equation}\label{Hdot1}
\frac{\dot{H}}{H^2}=-\frac{3}{2}\Big(1+\omega_D\Omega_D\Big).
\end{equation}
Also, if we take derivative of Eq. (\ref{HubbleHDE}) with respect
to time, after combining the result with (\ref{ConserveDEQ}) and
(\ref{Hdot1}), we obtain the EoS parameter as
\begin{equation}\label{EoS1}
\omega_D=-\frac{b^2}{c^2(1-c^2)}.
\end{equation}
The rate of deceleration/acceleration of the Universe expansion is
characterized by the deceleration parameter that encoded the
second derivative of the scale factor. In our case the
deceleration parameter is found to be
\begin{equation}\label{deceleration2}
q=-1-\frac{\dot{H}}{H^2}=\frac{1}{2}-\frac{3b^2}{2(1-c^2)}.
\end{equation}
It is clear that in the absence of the interaction where $b^2=0$,
we have $w_D=0$ and $q=1/2>0$, which implies that we encounter a
decelerated Universe.  Furthermore, in order to find more
sensitive discriminator of the expansion rate, we consider the
statefinder parameters which contain the third derivative of the
scale factor. The satefinder parameter $r$ can also be expressed
as
\begin{equation}\label{rr2}
r=1+3\frac{\dot{H}}{H^2}+\frac{\ddot{H}}{H^3}.
\end{equation}
Taking the time derivative of both sides Eq. (\ref{Friedeq1}) and
using Eqs. (\ref{r}), (\ref{ConserveCDMQ})-(\ref{rr2}), we can
obtain the statefinder pair $\{r,s\}$ as
\begin{eqnarray}\label{r3}
r&=&1+\frac{9b^2(c^2+b^2-1)}{2(1-c^2)^2},\\
s&=&\frac{c^2+b^2-1}{c^2-1}.
\end{eqnarray}
{The statefinder pair parameters $\{r, s\}$ become constant
depending on $b^2$ and $c^2$ parameters. For $c^2=1-b^2$ we have
$\{r,s\}=\{1,0\}$, which is exactly the result obtained for
spatially flat $\Lambda$CDM model \cite{Feng}. So we can find
relationship between the coupling constant $b^2$ and HDE constant
parameter $c^2$.}

Next, we consider the GO cutoff as the system's IR cutoff, namely
$L=(\alpha \dot{H}+\beta H^2)^{-1/2}$. This cutoff was first
introduced by Granda and Oliveros \cite{Granda} and can be
considered as the formal generalization of the Ricci scalar
curvature radius \cite{Gao}. The cosmological implications of the
HDE model with GO cutoff have been investigated in \cite{GO}. With
this IR cutoff, the energy density (\ref{HDE}) is written
\begin{equation}\label{GOHDE}
\rho_D=3M_p^2(\alpha H^2+\beta \dot{H}),
\end{equation}
where $\alpha$ and $\beta$ are constants which should be
constrained by observational data, and we have absorbed the constant
$c^2$ in $\alpha$ and $\beta$ in HDE density. For the case of GO
cutoff, we do not need to take into account the interaction
between DE and DM and without it we still have acceleration. Thus,
for economical reason, here we consider the noninteracting case with
$b^2=0=Q$, so the continuity equations read to
\begin{eqnarray}\label{cons1}
&& \dot{\rho}_m+3H\rho_m=0,\\
&& \dot{\rho}_D+3H(1+\omega_D)\rho_D=0.\label{cons2}
\end{eqnarray}
\begin{figure}[htp]
\begin{center}
\includegraphics[width=8cm]{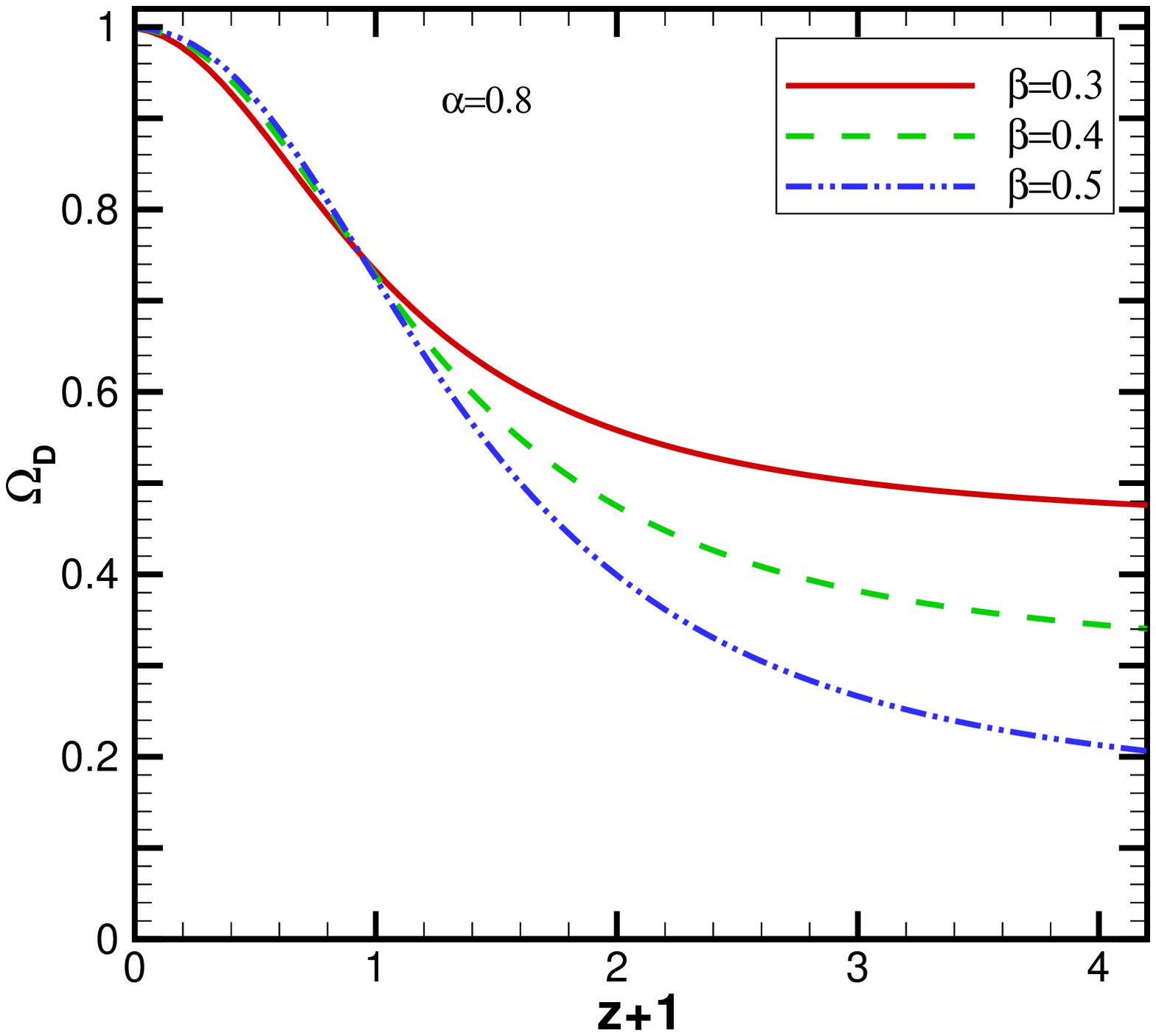}
\includegraphics[width=8cm]{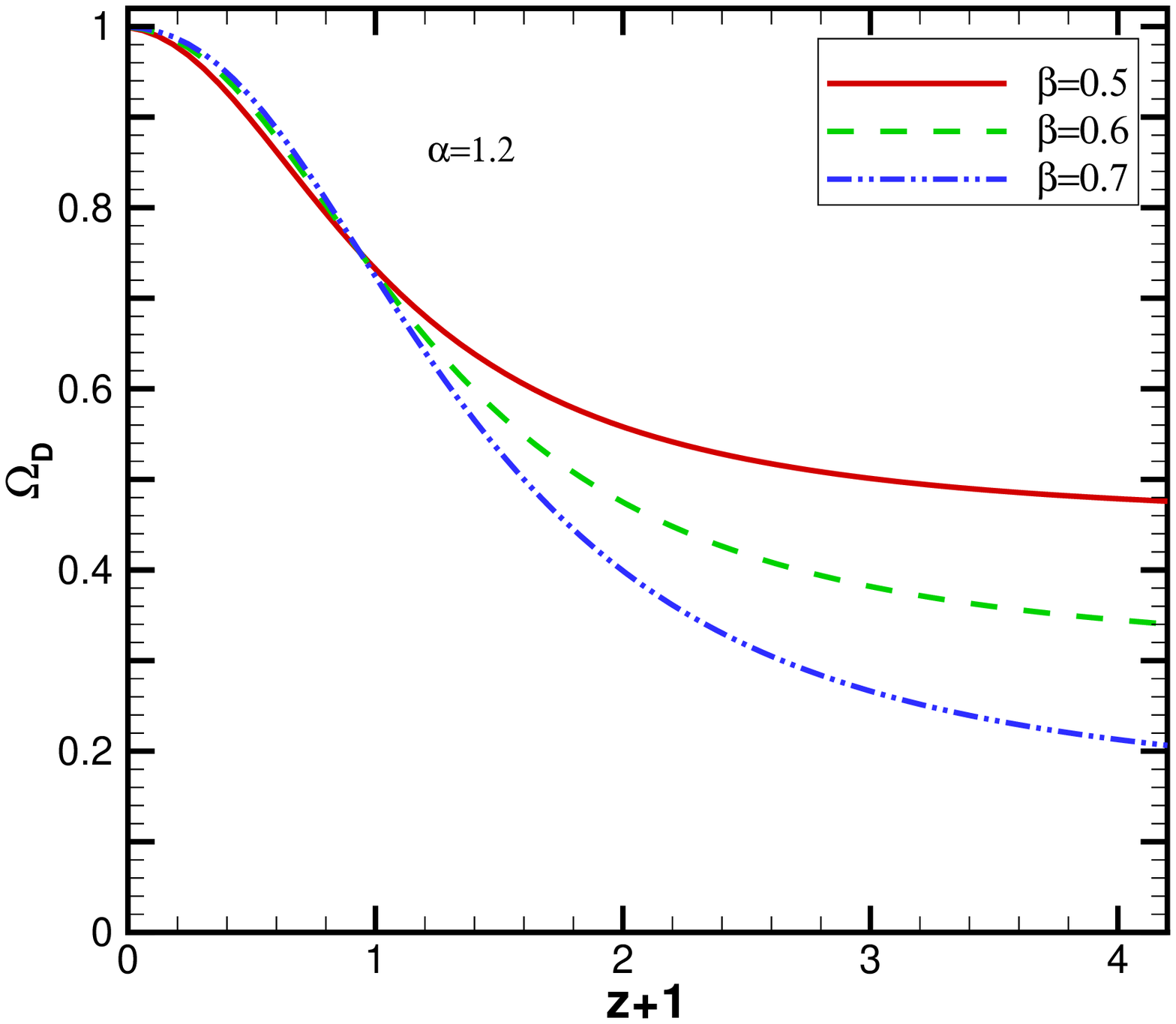}
\caption{The evolution of $\Omega_D$ versus redshift parameter $z$
for HDE with GO cutoff in standard cosmology. Left panel
corresponds to $\alpha<1$ and the right panel shows the case
$\alpha>1$.}\label{Omega}
\end{center}
\end{figure}
Now taking the time derivative of both sides of Eq.
(\ref{Friedeq1}) and using Eq. (\ref{Friedeq2}), we get
\begin{equation}\label{HDEdot}
\frac{\dot{\rho}_D}{3M_p^2H^3}=\frac{2\dot{H}}{H^2}+3(1-\Omega_D).
\end{equation}
From Eq. (\ref{GOHDE}) one can obtain
\begin{equation}\label{Friedeqdot}
\frac{\dot{H}}{H^2}=\frac{\Omega_D-\alpha}{\beta}.
\end{equation}
The equation of motion for the dimensionless HDE density is
obtained by taking the time derivative of relation
(\ref{Friedeq3})
\begin{equation}\label{Omegadot}
\dot{\Omega}_D=\frac{\dot{\rho}_D}{3M_p^2H^2}-2\Omega_D\frac{\dot{H}}{H}.
\end{equation}
Combining Eqs. (\ref{HDEdot}) and (\ref{Friedeqdot}) with
(\ref{Omegadot}), we arrive at
\begin{equation}\label{Omegadot2}
{\Omega}^\prime_D=(1-\Omega_D)
\left[3+\frac{2}{\beta}(\Omega_D-\alpha)\right].
\end{equation}
where a prime denotes derivative with respect to $x=\ln a$ and we
have used relation $\dot{\Omega}_D=H{\Omega}^\prime_D$. The
evolution of the dimensionless HDE density $\Omega_D$ in terms of
redshift parameter $1+z=a^{-1}$ is shown in Fig. (\ref{Omega}).
From these figures we see that at the early time
$\Omega_D\rightarrow 0$, while at the late time where
$a\rightarrow\infty$, the DE dominated, namely
$\Omega_D\rightarrow1$.

Combining Eqs. (\ref{GOHDE}), (\ref{HDEdot}) and
(\ref{Friedeqdot}) with (\ref{cons2}), we can obtain the EoS
parameter as
\begin{equation}\label{EoS2}
\omega_D=-\frac{1}{\Omega_D}\left[1+\frac{2}{3}\left(\frac{\Omega_D-\alpha}{\beta}\right)\right]
\end{equation}
The deceleration parameter can be expressed in this model as
\begin{equation}\label{q2}
q=-1-\frac{\dot{H}}{H^2}=-1-\frac{\Omega_D-\alpha}{\beta}.
\end{equation}
\begin{figure}[htp]
\begin{center}
\includegraphics[width=8cm]{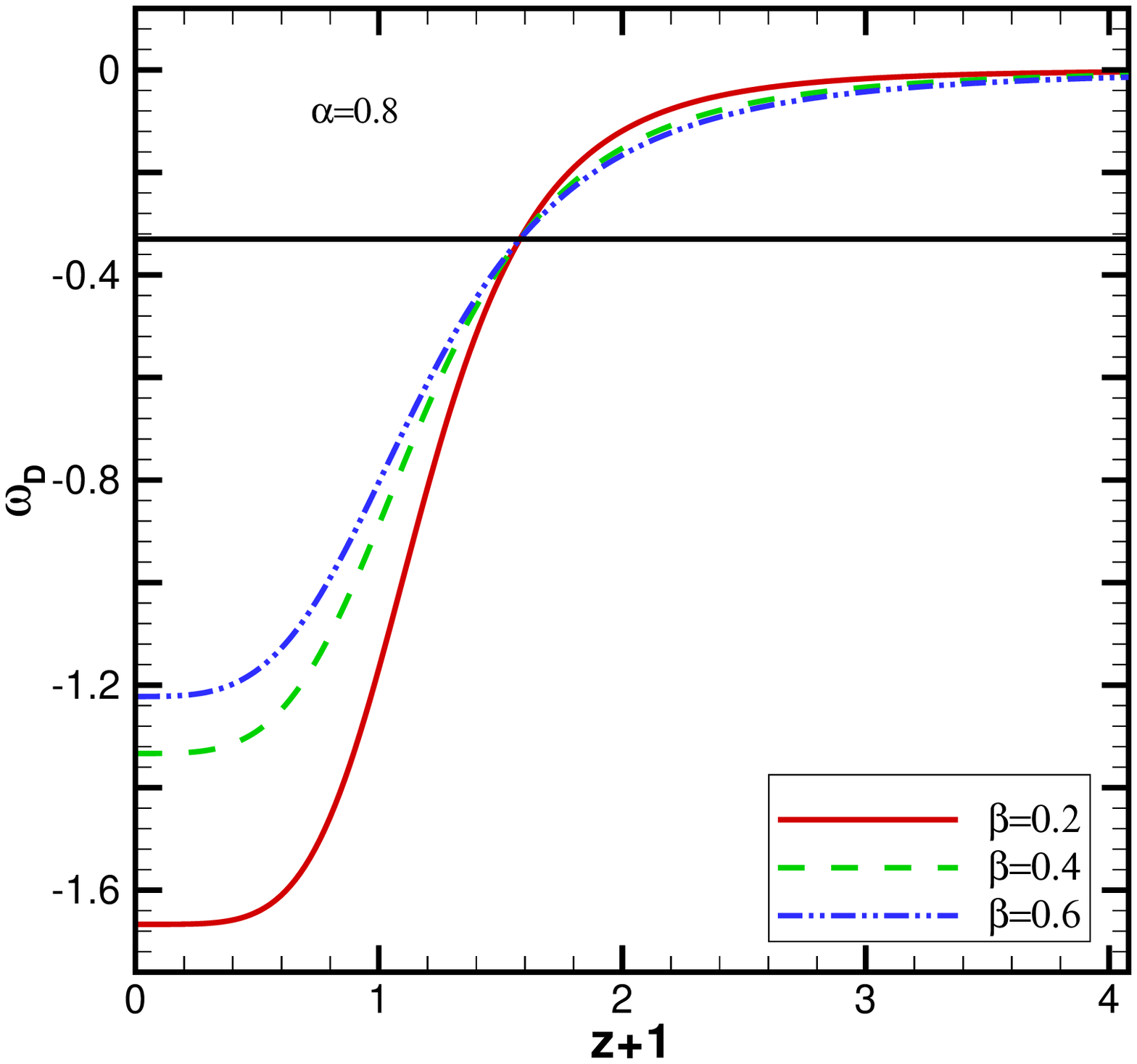}
\includegraphics[width=8cm]{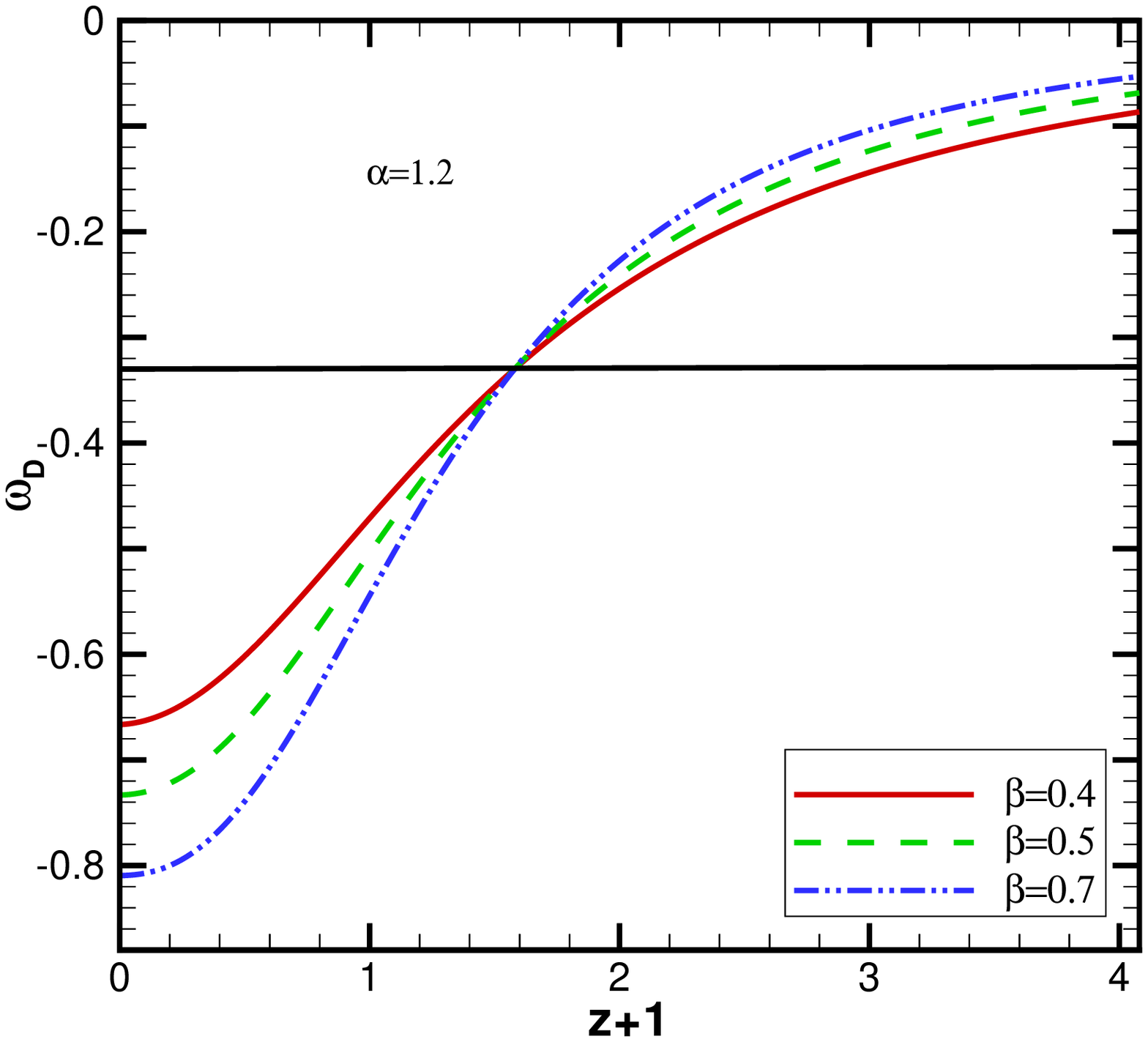}
\caption{The evolution of  $\omega_D$ versus redshift parameter
$z$ for HDE with GO cutoff in standard cosmology. Left panel
corresponds to $\alpha<1$ and the right panel shows the case
$\alpha>1$.}\label{EoS1}
\end{center}
\end{figure}
\begin{figure}[htp]
\begin{center}
\includegraphics[width=8cm]{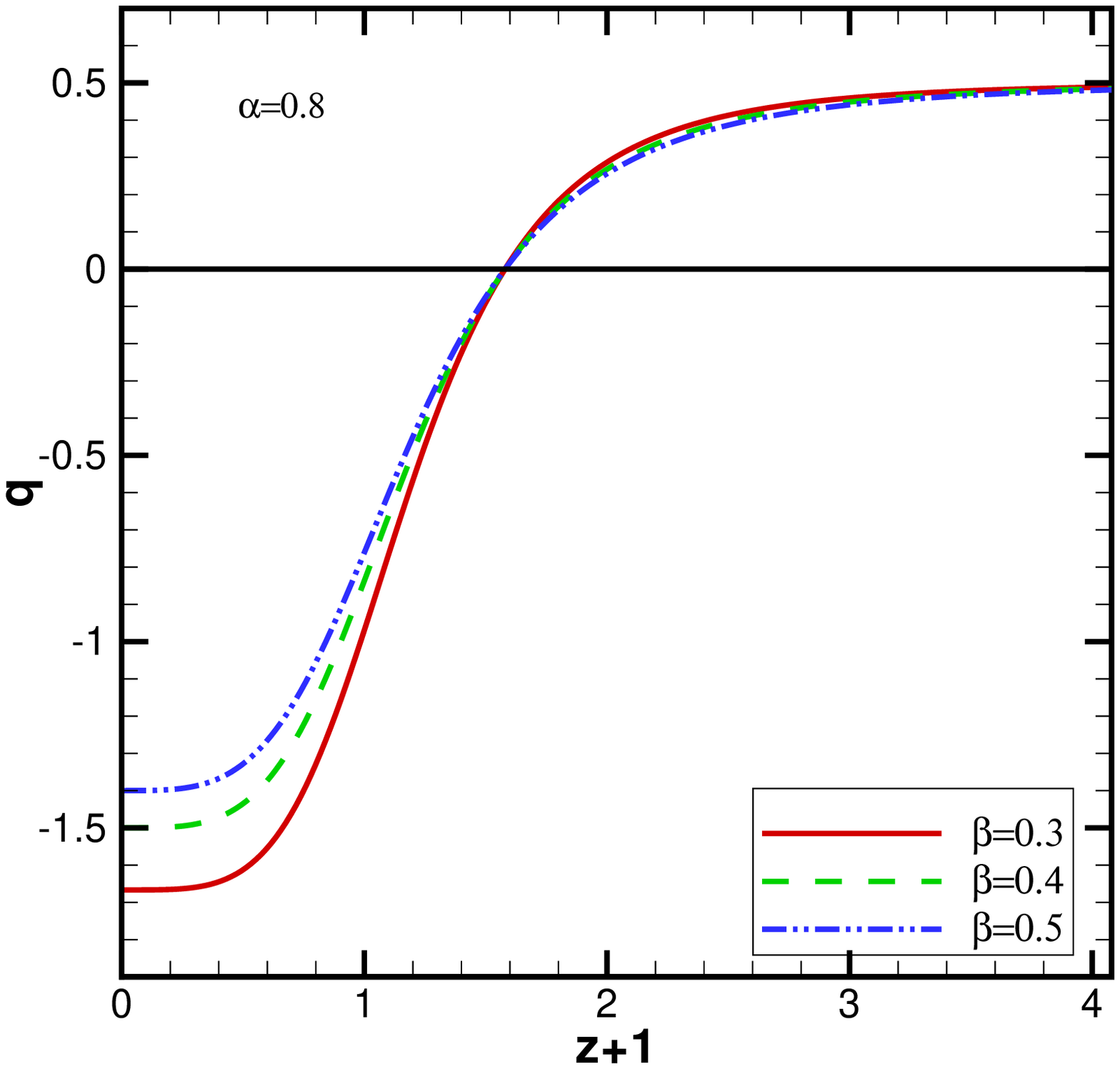}
\includegraphics[width=8cm]{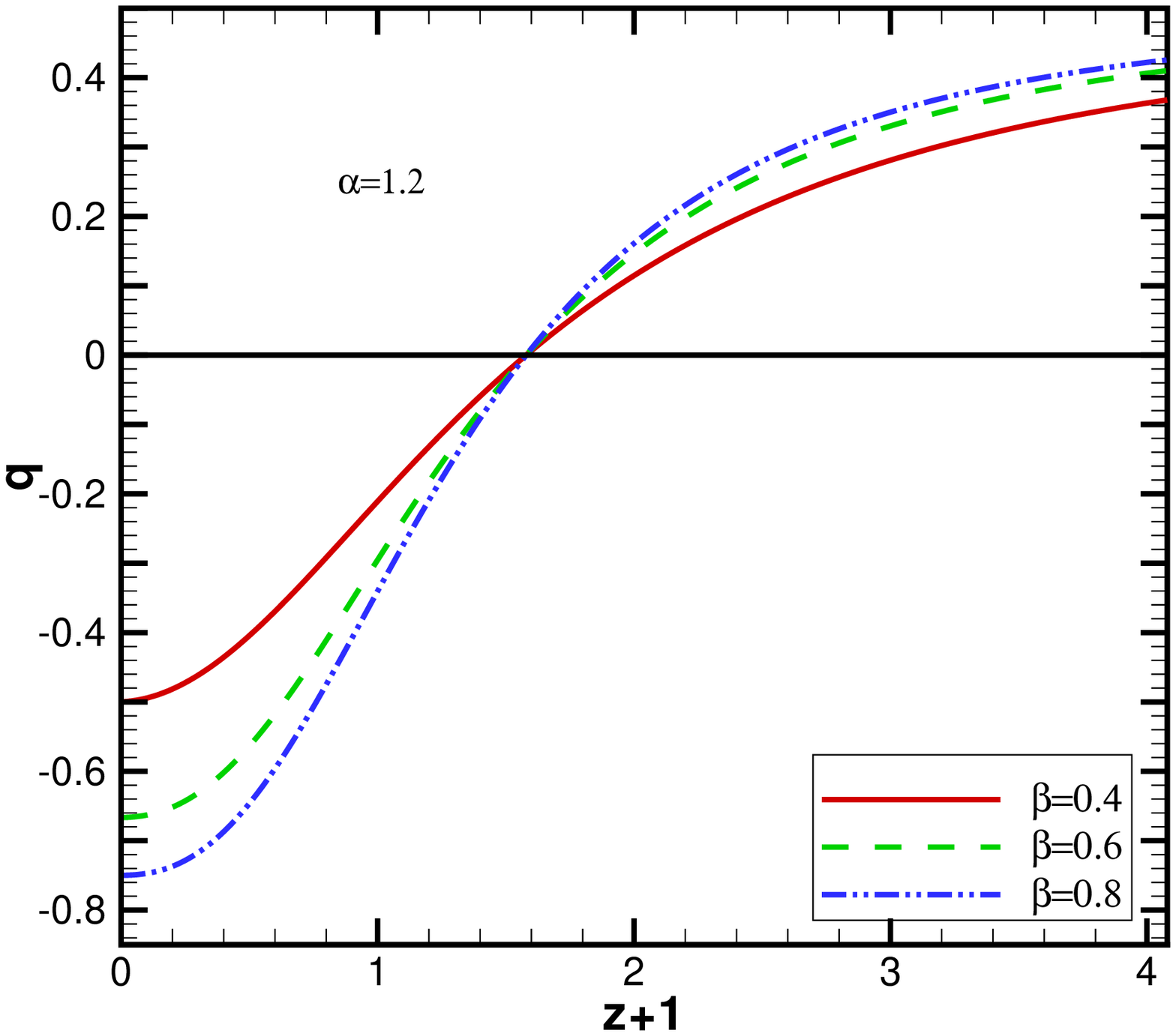}
\caption{The evolution of the  deceleration parameter $q$ against
redshift parameter $z$ for HDE with GO cutoff in standard
cosmology.}\label{q1}
\end{center}
\end{figure}
When $\Omega_D\rightarrow\alpha=1$ we have $\omega_D=-1$ and
$q=-1$, similar to the cosmological constant. Therefore, we
investigate the behavior of $\omega_D$ for values close to
$\alpha=1$. We consider the case with $\alpha<1$ and $\alpha>1$
separately. We find that for both cases, our Universe has a
transition from deceleration to the acceleration phase around
$z\approx 0.6$. The behavior of $\omega_D$ and $q$ are plotted in
figures 2 and 3. From these figures we see that for $\alpha<1$ the
EoS parameter can cross the phantom line in the future, i.e.,
$\omega_D<-1$,
while for $\alpha>1$ the EoS parameter is always larger than $-1$.\\
Taking the time derivative of Eq. (\ref{Friedeqdot}), after using
(\ref{Omegadot2}), we obtain
\begin{equation}\label{H3}
\frac{\ddot{H}}{H^3}=\frac{2}{\beta^2}(\Omega_D-\alpha)(1-\alpha)+\frac{3}{\beta}(1-\Omega_D).
\end{equation}
Inserting Eqs. (\ref{Friedeqdot}) and (\ref{H3}) in (\ref{rr2}),
after some calculations, we obtain statefinder pair parameters
$\{r,s\}$ as follow,
\begin{eqnarray}
r&=&1+\frac{(1-\alpha)}{\beta}\left[3+\frac{2}{\beta}\left(\Omega_D-\alpha\right)\right],\\
s&=&\frac{\alpha-1}{6\beta}.
\end{eqnarray}
It can be easily seen that for $\alpha=1$, we have $\{r=1,s=0\}$
as expected. The statefinder parameter $r(z)$ is plotted in figure
4 and 5 for two cases. These results are compatible
 with the arguments given in \cite{Alam, Sahni}.
\begin{figure}[htp]
\begin{center}
\includegraphics[width=8cm]{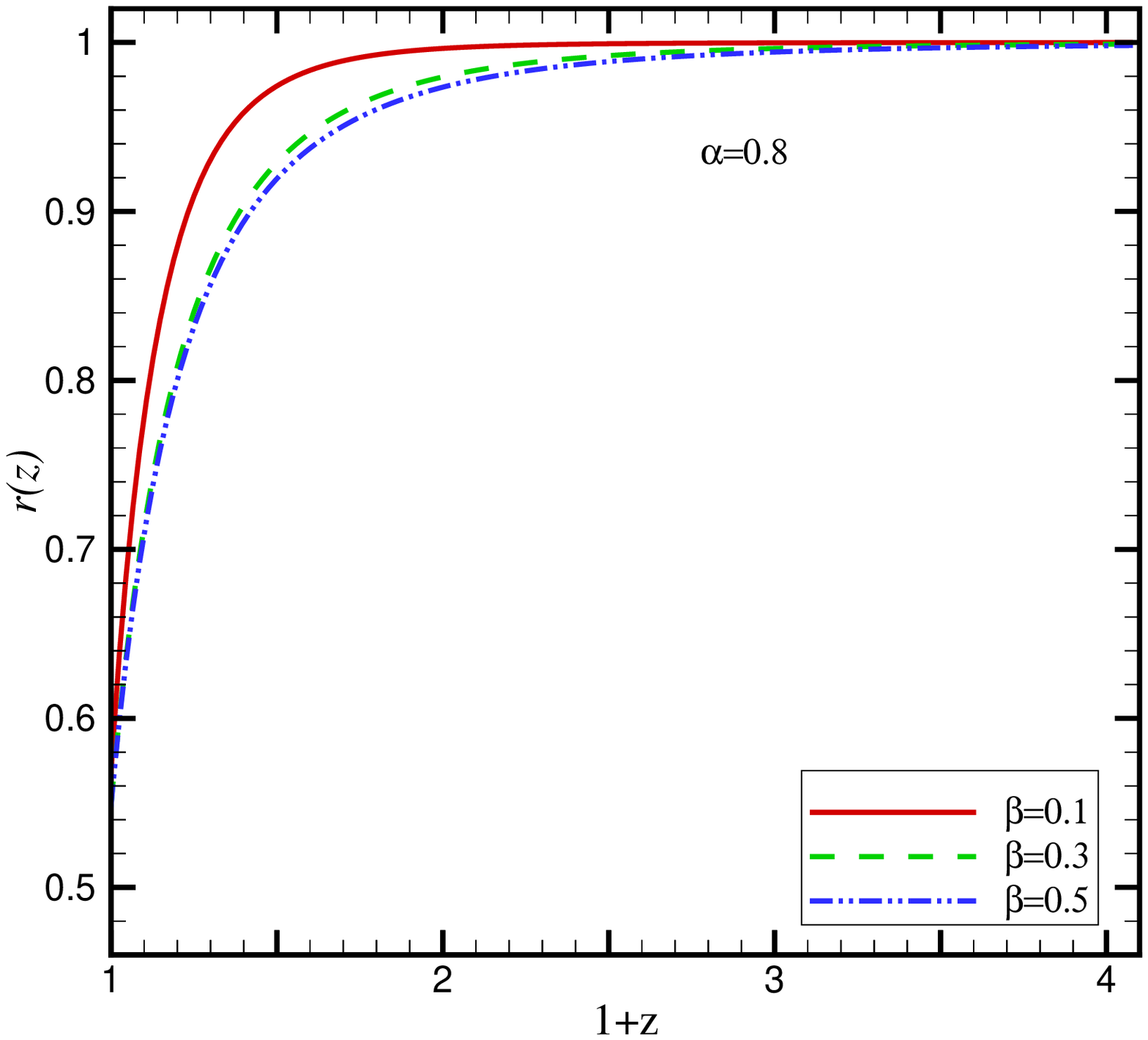}
\includegraphics[width=8cm]{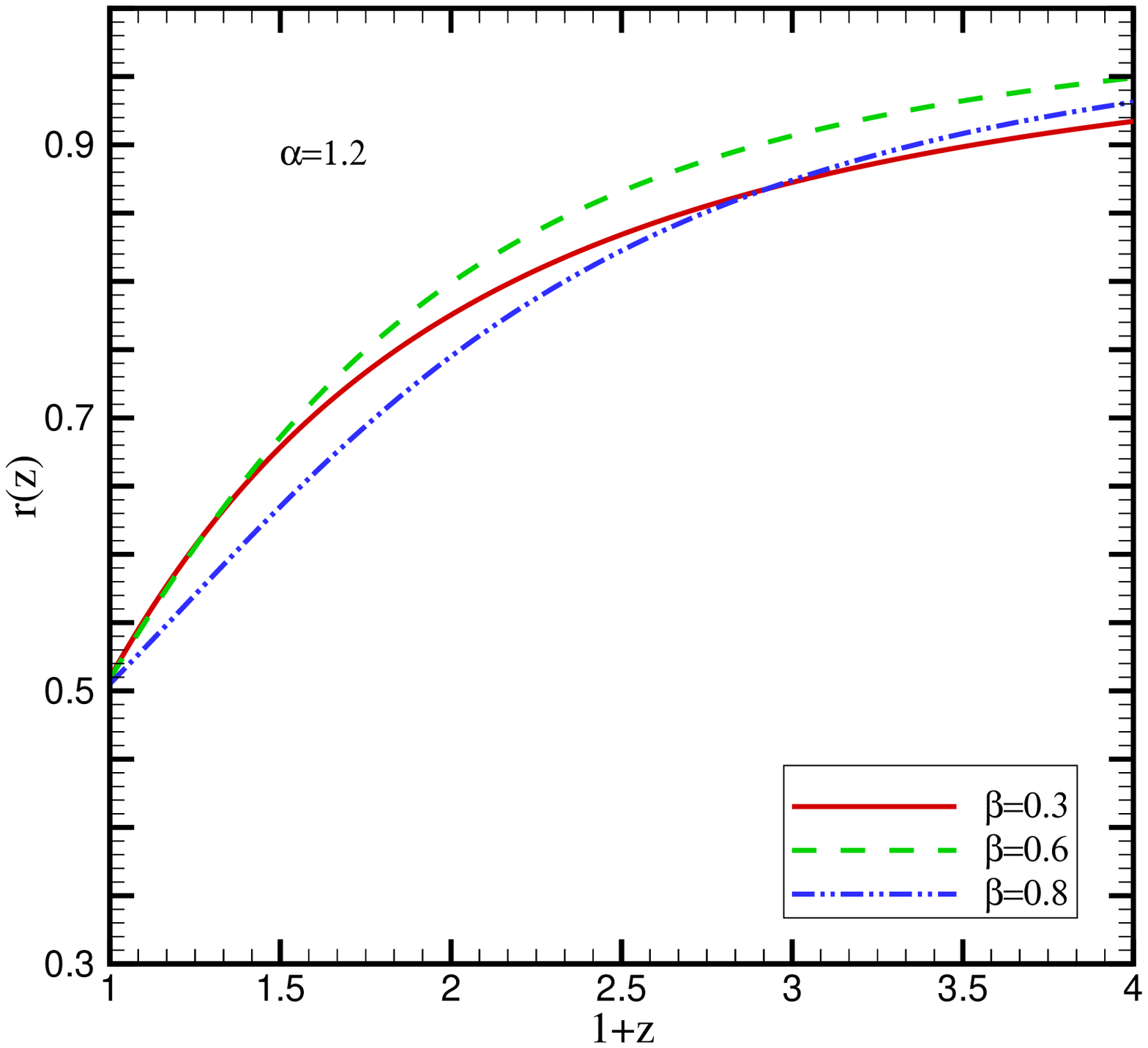}
\caption{The evolution of the statefinder parameter $r$ versus
redshift parameter $z$ for $\alpha<1$ (left panel) and $\alpha>1$
(right panel).}\label{r1}
\end{center}
\end{figure}
\begin{figure}[htp]
\begin{center}
\includegraphics[width=8cm]{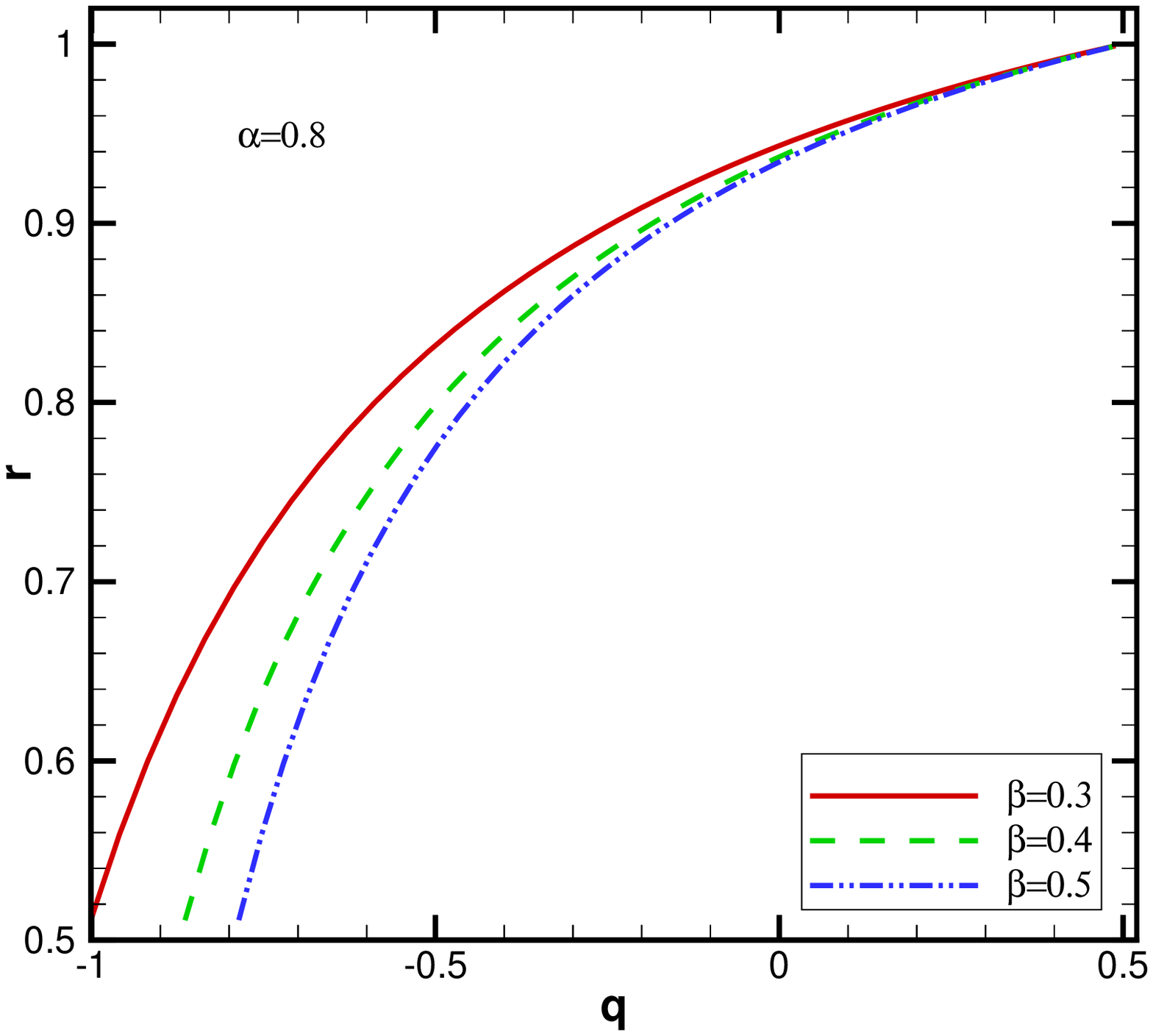}
\includegraphics[width=8cm]{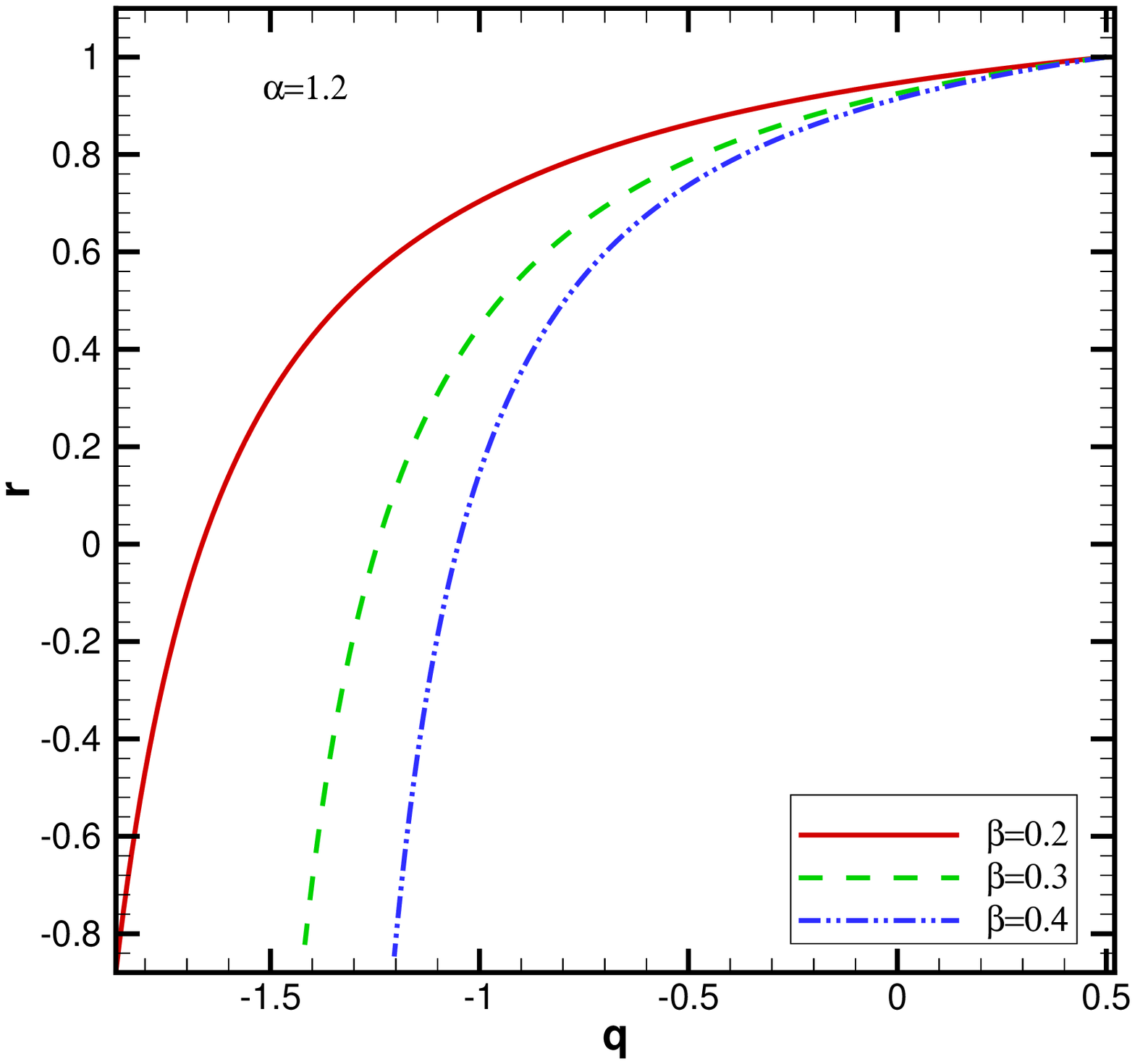}
\caption{The evolution of the statefinder parameter $r$ versus
declaration parameter $q$ for $\alpha<1$ (left panel) and
$\alpha>1$ (right panel).}\label{r1}
\end{center}
\end{figure}
%%%%%%%%%%%%%%%%%%%%%%%%%%%%%%%%%%%%%%%%%%%%%%%%%%%%%%%%%%%%%%%%%%%%%%%%%%%%%%%%%%%%%%
\section{DGP bareworld model}
In this section, we would like to extend the study to braneworld
scenario. According to this scenario, our Universe is realized as
a 3-brane embedded in a five dimensional spacetime. A special
version of braneworld scenario was proposed by Dvali-
Gabadadze-Porrati (DGP) \cite{Dvali}, in which our
four-dimensional FRW Universe embedded in a five-dimensional
Minkowski bulk with infinite size. The recovery of the usual
gravitational laws in this picture is obtained by adding to the
action of the brane an Einstein-Hilbert term computed with the
brane intrinsic curvature. The presence of such a term in the
action is generically induced by quantum corrections coming from
the bulk gravity and its coupling with matter living on the brane.
It was argued that this term should be included in a large class
of theories for self-consistency \cite{Dvali2,Alder}. In this
model the cosmological evolution on the brane is described by an
effective Friedmann equation which incorporate the non-trivial
bulk effects onto the brane. The modified Friedmann equation in
this model is given by  \cite{Def}
\begin{equation}\label{Friedeq4}
H^2+\frac{k}{a^2}=\Big(\sqrt{\frac{\rho}{3M_{\rm
pl}^2}+\frac{1}{4r^2_c}}+\frac{\epsilon}{2r_c}\Big)^2,
\end{equation}
where $\rho=\rho_m+\rho_D$ is the total cosmic fluid energy
density on the brane. The cross over length $r_c$ is given by
\cite{Def}
\begin{equation}
r_c=\frac{M_{\rm pl}^2}{2M_{5}^3}=\frac{G_5}{2G_4},
\end{equation}
and is defined as a scale which sets a length beyond which gravity
starts to leak out into the bulk \cite{Def}. In other words, it is
the distance scale reflecting the competition between 4D and 5D
effects of gravity. The parameter $\epsilon=\pm1$ represents the
two branches of solutions of the DGP model \cite{Def}. For
$\epsilon=+1$ branch, there is an accelerated expansion at the
late time without invoking any kind of DE or other components of
energy \cite{Gaffari}, however, it suffers the ghost instabilities
problem \cite{Koyama}. Besides, it cannot realize phantom divide
crossing by itself. On the other hand, for $\epsilon=-1$ branch
cannot undergos an accelerated expansion phase without additional
DE component\cite{Deffayet1}. For $H^{-1}\ll r_c$ (early time),
the Friedmann equation in standard cosmology is recovered,
\begin{equation}
H^2+\frac{k}{a^2}=\frac{\rho}{3 M_{\rm pl}^2}.
\end{equation}
For the spatially flat DGP braneworld ($k=0$), the Friedmann
equation (\ref{Friedeq4}) reduces to
\begin{equation}\label{Friedeq5}
H^2-\frac{\epsilon}{r_c}H=\frac{1}{3M_p^2}(\rho_m+\rho_D).
\end{equation}
We can rewrite this equation as
\begin{equation}\label{Friedeq6}
1-2\epsilon\sqrt{\Omega_{r_c}}=\Omega_m+\Omega_D,
\end{equation}
where we have defined
\begin{equation}\label{Friedeq7}
\Omega_{r_c}=\frac{1}{4H^2r_c^2}.
\end{equation}
In the remaining part of this paper, we shall assume the
continuity equations hold on the brane,
\begin{eqnarray}\label{ConserveCDM2}
&&\dot{\rho}_m+3H\rho_m=0,\\
&&\dot{\rho}_D+3H(1+\omega_D)\rho_D=0. \label{ConserveDE2}
\end{eqnarray}
Following the previous section, we shall consider two cutoffs,
namely, the Hubble radius and GO cutoff, as systems's IR cutoff in
the HDE density. First we consider $L=H^{-1}$. Inserting
$\rho_D=3c^2M_p^2H^2$ into the Friedmann equation
(\ref{Friedeq5}), we get
\begin{equation}\label{Friedeq8}
H^2(1-c^2)-\frac{\epsilon}{r_c}H=\frac{\rho_m}{3M_p^2},
\end{equation}
which can also be written as
\begin{equation}\label{Fr66}
1-c^2-2\epsilon\sqrt{\Omega_{r_c}}=\Omega_m.
\end{equation}
Taking the time derivative of Eq. (\ref{Friedeq8}) after using
Eqs. (\ref{Friedeq7}) and (\ref{ConserveCDM2}), we arrive at
\begin{equation}\label{Friedeqdot2}
2\frac{\dot{H}}{H^2}=\frac{-3\Omega_m}{1-c^2-\epsilon\sqrt{\Omega_{r_c}}}.
\end{equation}
We can also obtain the equation of motion for $\Omega_{r_c}$. For
this purpose, we first take the time derivative of Eq.
(\ref{Friedeq7}), and then combining the result with Eqs.
(\ref{Friedeq6}) and (\ref{Friedeqdot2}), we arrive at
\begin{equation}
{\Omega}^{\prime}_{r_c}=\frac{-3\Omega_{r_c}(1-c^2-2\epsilon\sqrt{\Omega_{r_c}})}{1-c^2-\epsilon\sqrt{\Omega_{r_c}}}.
\end{equation}
From Eq. (\ref{ConserveDE2}) we have
\begin{equation}
\omega_D=-\frac{\dot{\rho}_D}{3H\rho_D}-1,
\end{equation}
\begin{figure}[htp]
\begin{center}
\includegraphics[width=8cm]{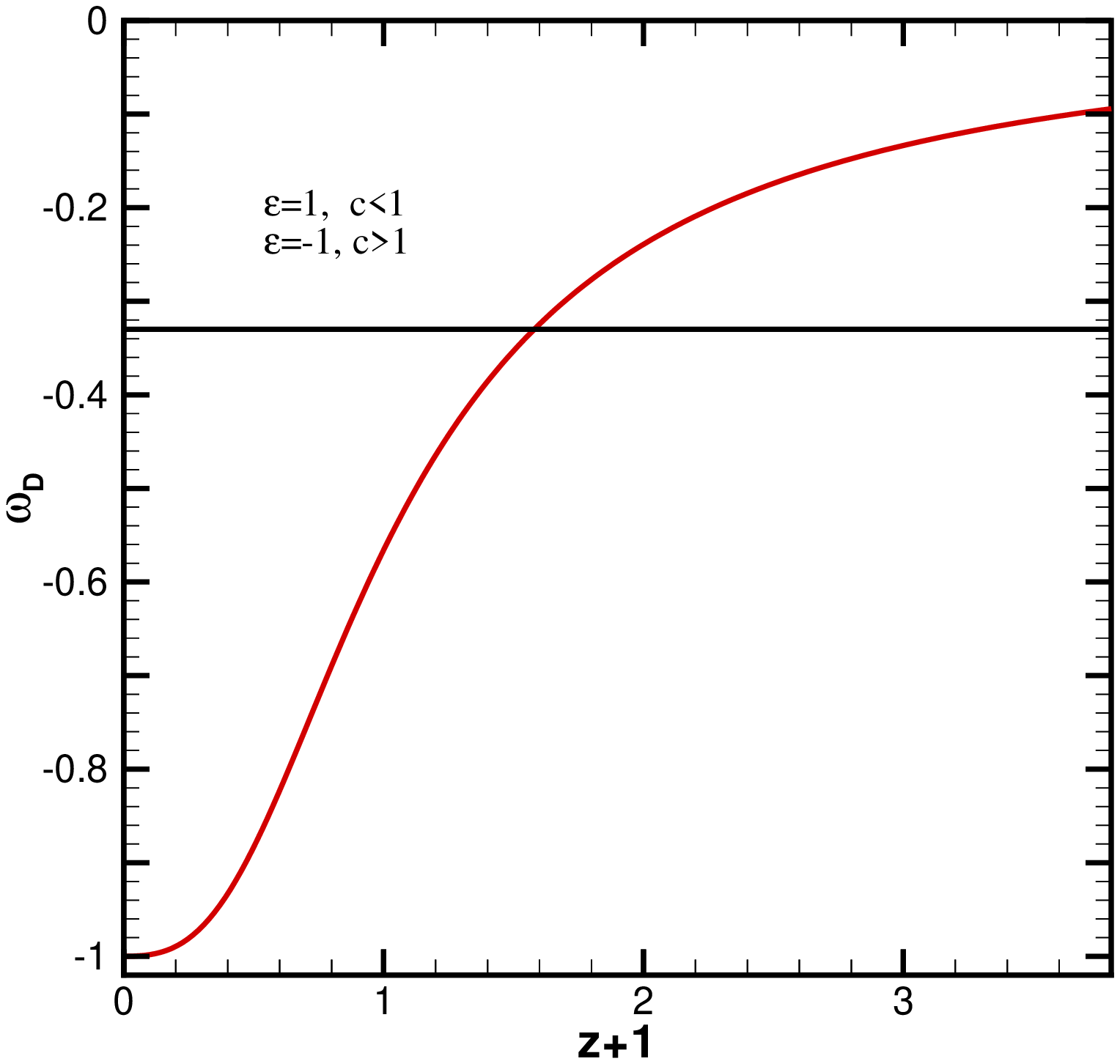}
\includegraphics[width=8cm]{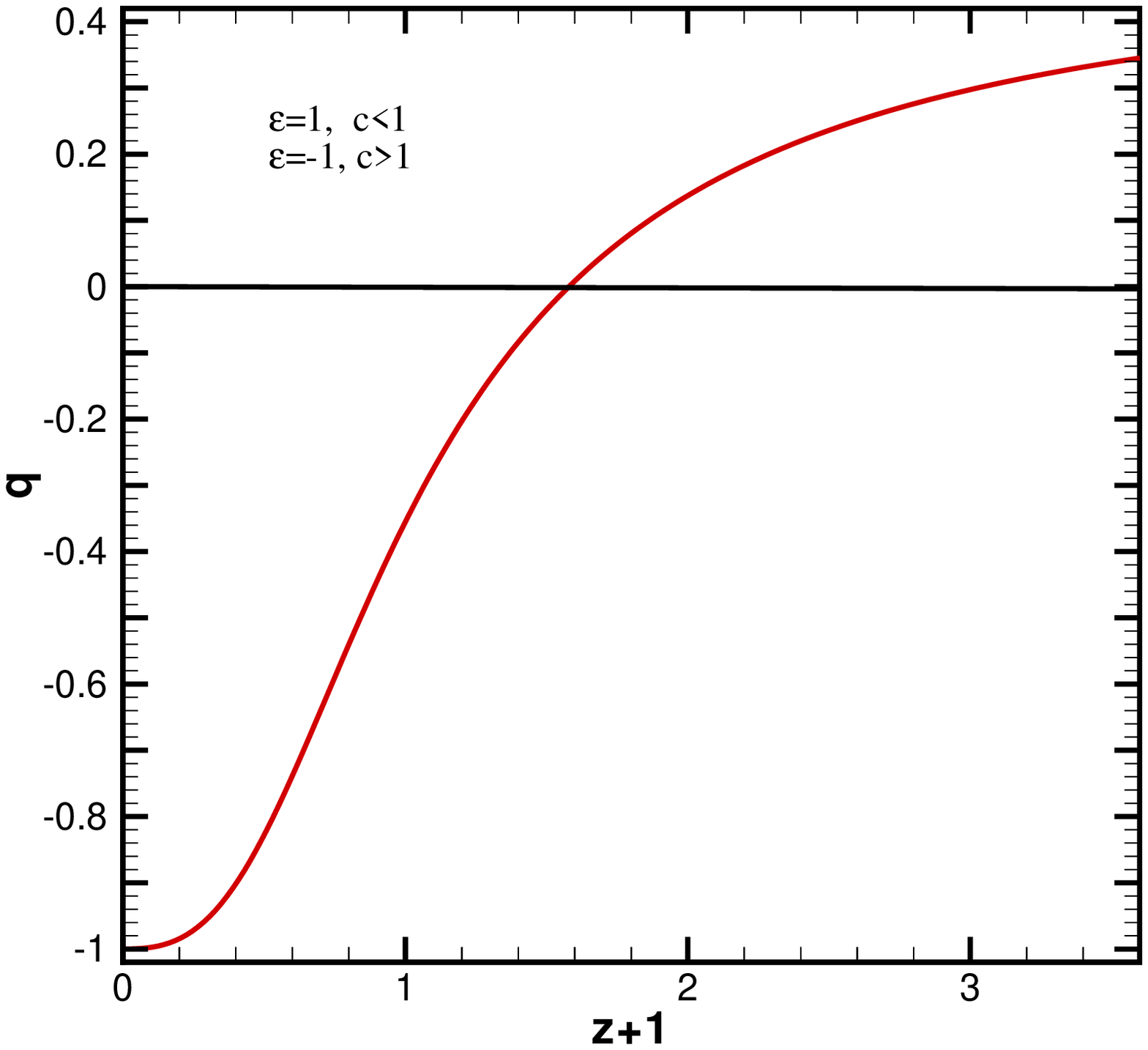}
\caption{The evolution of  $w_{D}$ and the deceleration parameter
$q$ versus redshift parameter $z$ for the two branches of DGP
braneworld. For both cases $(\epsilon=1, c<1)$ and $(\epsilon=-1,
c>1)$, the diagrams are quite similar to each other.}\label{w5}
\end{center}
\end{figure}
while taking the time derivative of $\rho_D=3c^2M_p^2H^2$, we find
\begin{equation}
\dot{\rho}_D=2\rho_D\frac{\dot{H}}{H},
\end{equation}
and hence the EoS parameter reads to
\begin{equation}\label{wd}
\omega_D=-\frac{2}{3}\frac{\dot{H}}{H^2}-1=\frac{-\epsilon
\sqrt{\Omega_{r_c}}}{1-c^2-\epsilon \sqrt{\Omega_{r_c}}},
\end{equation}
where in the last step, we have used (\ref{Fr66}) and
(\ref{Friedeqdot2}). For $r_c\gg1$ ($\Omega_{r_c}\rightarrow 0$),
the effects of the extra dimension vanishes and hence we reach to
the standard cosmology regime. In this case $\omega_D\rightarrow
0$, which is an expected result. This is due to the fact that in the absence of the
interaction, the choice of $L=H^{-1}$ leads to a wrong EoS, namely
that of dust \cite{Li}, and hence this choice for the IR cutoff
in standard cosmology cannot leads to the accelerated expansion.
However, from (\ref{wd}) we see that in the DGP braneworld, the
natural choice for the IR cutoff in flat Universe, namely the
Hubble radius $L=H^{-1}$, can produce the accelerated expansion.
This is one of the interesting results we find in this paper.

The deceleration parameter is also obtained as
\begin{equation}\label{qdgp}
q=-1-\frac{\dot{H}}{H^2}=\frac{1}{2}-\frac{3\epsilon\sqrt{\Omega_{r_c}}}{2(1-c^2-\epsilon\sqrt{\Omega_{r_c}})}.
\end{equation}
The evolution of $\omega_D$ and $q$ are plotted numerically in
figure \ref{w5}, where we consider two branches of DGP braneworld
$\epsilon=1$ and $\epsilon=-1$ with different values of $c$. In
order to have reasonable behaviour for the evolution of the DE
during the history of the Universe, we should take $c<1$ in case
of $\epsilon=1$, while  $c>1$ in case of $\epsilon=-1$. {For both
cases diagrams have similar behavior.} From figure \ref{w5}, we
find that for $\epsilon=\pm 1$ we have a transition from
deceleration to acceleration phase around $z\approx0.6$. From
figure \ref{w5} one can see that EoS parameter $\omega_D$
gradually tend to $-1$ with the evolution of the Universe, while
it is larger than $-1$ throughout the evolution of Universe, which
is an indication for quintessence behavior \cite{Cai2}.

Taking the time derivative of Eq. (\ref{Friedeqdot2}) and using
Eqs. (\ref{Friedeq7}) and (\ref{ConserveCDM2}), we arrive at
\begin{equation}\label{Friedeqdot3}
\frac{\ddot{H}}{H^3}=-\Big(\frac{\dot{H}}{H^2}\Big)^2\frac{\epsilon\sqrt{\Omega_{r_c}}}{(1-c^2-\epsilon\sqrt{\Omega_{r_c}})}
+\frac{9(1-c^2-2\epsilon\sqrt{\Omega_{r_c}})}{2(1-c^2-\epsilon\sqrt{\Omega_{r_c}})}.
\end{equation}
Using Eqs. (\ref{Friedeqdot2}), (\ref{qdgp}) and
(\ref{Friedeqdot3}), after some calculations, we obtain the
statefinder pair parameters as
\begin{eqnarray}
r&=&1+3\frac{\dot{H}}{H^2}+\frac{\ddot{H}}{H^3}=1-\frac{9\epsilon\sqrt{\Omega_{r_c}}(1-c^2-2\epsilon\sqrt{\Omega_{r_c}})^2}{4(1-c^2-\epsilon\sqrt{\Omega_{r_c}})^3},\\
s&=&\frac{r-1}{3(q-1/2)}=\frac{(1-c^2-2\epsilon\sqrt{\Omega_{r_c}})^2}{2(1-c^2-\epsilon\sqrt{\Omega_{r_c}})^2}.
\end{eqnarray}

Next, we choose the GO cutoff for the HDE density in the framework
of the DGP braneworld. We restrict our study to the current
cosmological epoch, and hence we are not considering the
contributions from matter and radiation by assuming that the DE
dominates, thus the Friedman equation becomes simpler.
Substituting (\ref{GOHDE}) into Friedmann equation
(\ref{Friedeq5}), we can obtain the differential equation for the
Hubble parameter as
\begin{equation}\label{Friedeq9}
H^2(1-\alpha)-\frac{\epsilon}{r_c}H-\beta \dot{H}=0.
\end{equation}
Solving this equation, the Hubble parameter is obtained as
\begin{equation}\label{Hubble1}
H(t)=\frac{\epsilon}{r_{\rm c}(1-\alpha)+c_1\epsilon
e^{\frac{\epsilon t}{\beta r_{\rm c}}}},
\end{equation}
where $c_1$ is a constant of integration. Since for $r_{\rm
c}\gg1$, the effects of the extra dimension should be disappeared
and the result of \cite{Granda}, namely
\begin{equation}\label{Hubble2}
H(t)=\frac{\beta}{\alpha-1}\frac{1}{t},
\end{equation}
must be restored, thus the constant $c_1$ should be chosen as
\cite{Gaffari}
\begin{equation}
c_1=\frac{r_{\rm c}(\alpha-1)}{\epsilon}.
\end{equation}
Substituting $c_1$ in Eq. (\ref{Hubble1}), we obtain
\begin{equation}\label{Hubble3}
H(t)=\frac{\epsilon}{r_{\rm c}(1-\alpha)(1-e^{\frac{\epsilon
t}{\beta r_{c}}})}.
\end{equation}
In term of the redshift parameter, we have \cite{Gaffari}
\begin{eqnarray}\label{Hubblez}
H(z)&=&\frac{\epsilon}{r_{\rm
c}(1-\alpha)}\left[1+(1+z)^{\frac{\alpha-1}{\beta}}\right].
\end{eqnarray}
Solving (\ref{Hubble3}) for the scale factor, yields
\begin{equation}\label{scale}
a(t)=a_0 \Big(e^{\frac{-\epsilon t}{\beta r_c}}-1\Big)^{\frac{\beta}{\alpha-1}}.
\end{equation}
Taking the time derivative of Eq. (\ref{Hubble3}), we arrive at
\begin{eqnarray}\label{Hdot2}
\dot{H}&=&\frac{e^{\frac{\epsilon t}{\beta r_c}}}{\beta
r_c^2(1-\alpha)(1-e^{\frac{\epsilon t}{\beta r_c}})^2},\\
\ddot{H}&=&\frac{\epsilon e^{\frac{\epsilon t}{\beta
r_c}}}{\beta^2r_c^3(1-\alpha)(1-e^{\frac{\epsilon t}{\beta
r_c}})^2} \Big(1+\frac{2e^{\frac{\epsilon t}{\beta
r_c}}}{1-e^{\frac{\epsilon t}{\beta r_c}}}\Big). \label{Hdot3}
\end{eqnarray}
Therefore, it is easy to show that
\begin{eqnarray}\label{H4}
\frac{\dot{H}}{H^2}&=&\frac{1-\alpha}{\beta}e^{\frac{\epsilon
t}{\beta r_c}}\\
\frac{\ddot{H}}{H^3}&=&\frac{(1-\alpha)^2}{\beta^2}e^{\frac{\epsilon
t}{\beta r_c}} \Big(1+e^{\frac{\epsilon t}{\beta
r_c}}\Big)\label{H5}
\end{eqnarray}
In this case the EoS and the deceleration parameters are obtained
as \cite{Gaffari}
\begin{equation}\label{wDz}
\omega_D(z)=-1-\frac{1-\alpha}{3\beta}
\Big[\frac{(1+\alpha)(1+z)^{\frac{1-\alpha}{\beta}}+2}
{(1+(1+z)^{\frac{1-\alpha}{\beta}})(1+\alpha(1+z)^{\frac{1-\alpha}{\beta}})}\Big],
\end{equation}
\begin{equation}\label{q1}
q(z)=-1-\frac{1-\alpha}{\beta\left[1+(1+z)^\frac{1-\alpha}{\beta}\right]}.
\end{equation}
\begin{figure}[htp]
\begin{center}
\includegraphics[width=8cm]{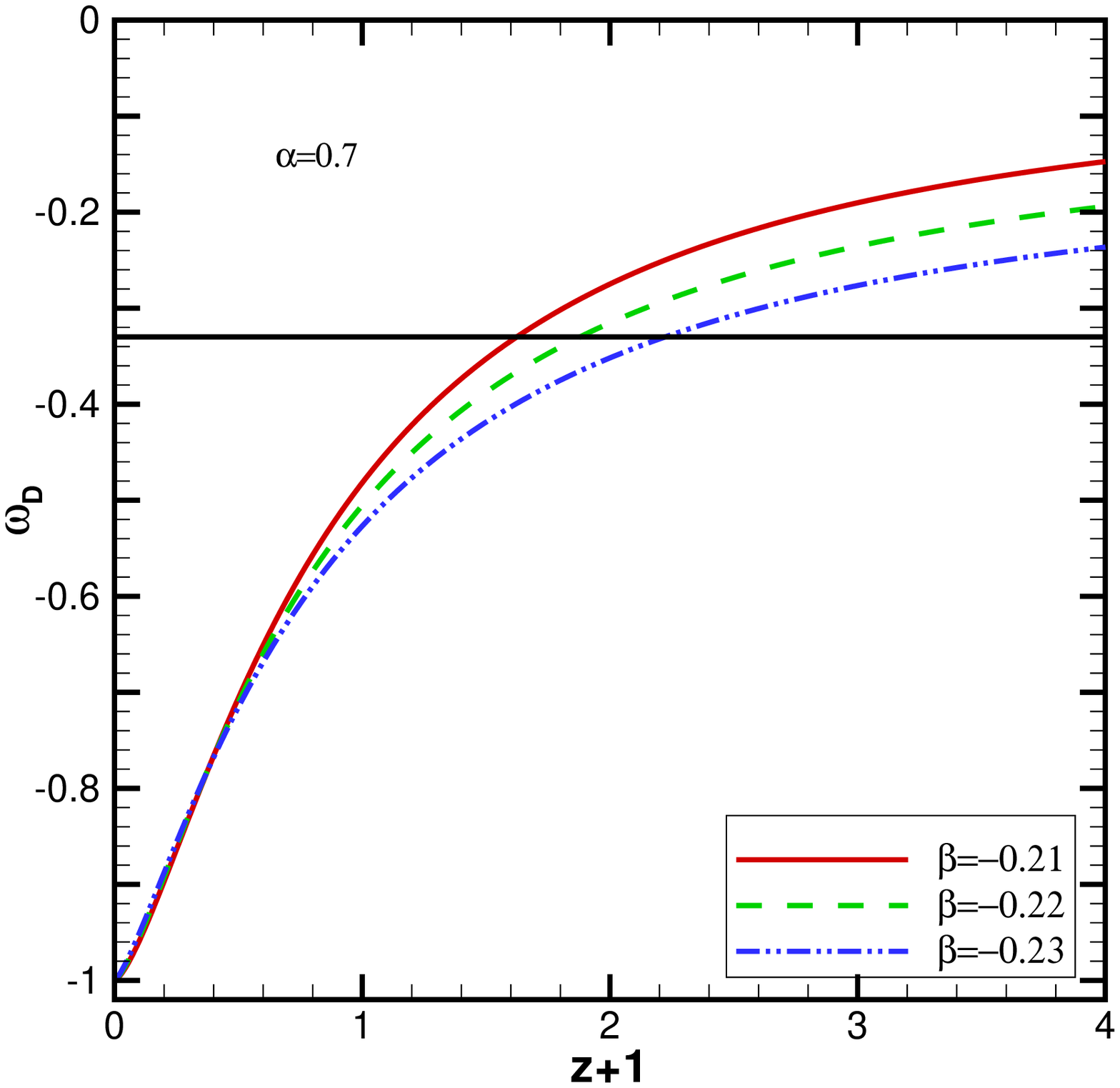}
\includegraphics[width=8cm]{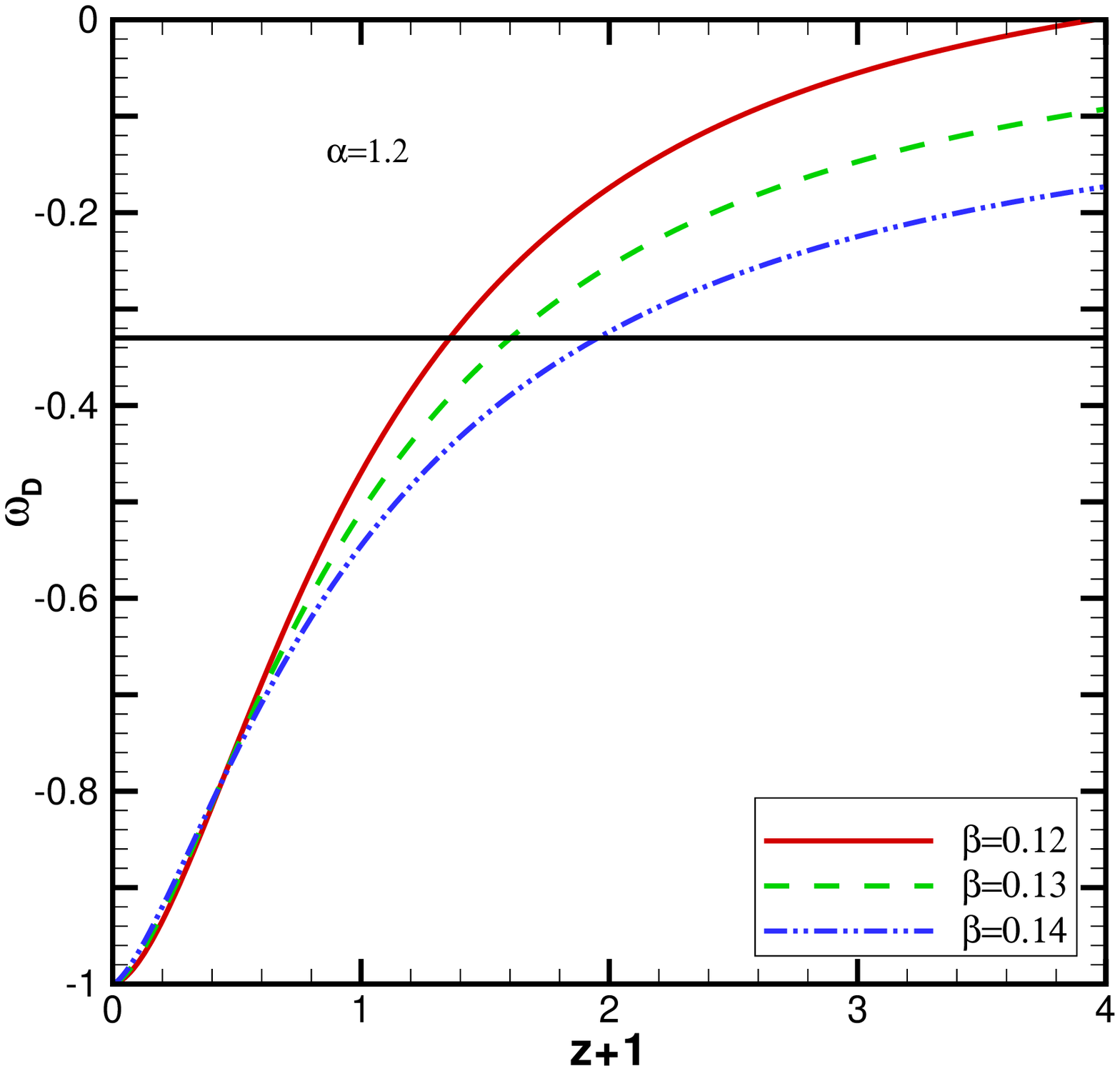}
\caption{The evolution of  $w_{D}$ versus redshift parameter $z$
for HDE with GO cutoff in DGP braneworld. Left panel corresponds
to $\alpha<1$ and $\beta<0$, while the right panel shows the case
$\alpha>1$ and $\beta>0$. }\label{w3}
\end{center}
\end{figure}

\begin{figure}[htp]
\begin{center}
\includegraphics[width=8cm]{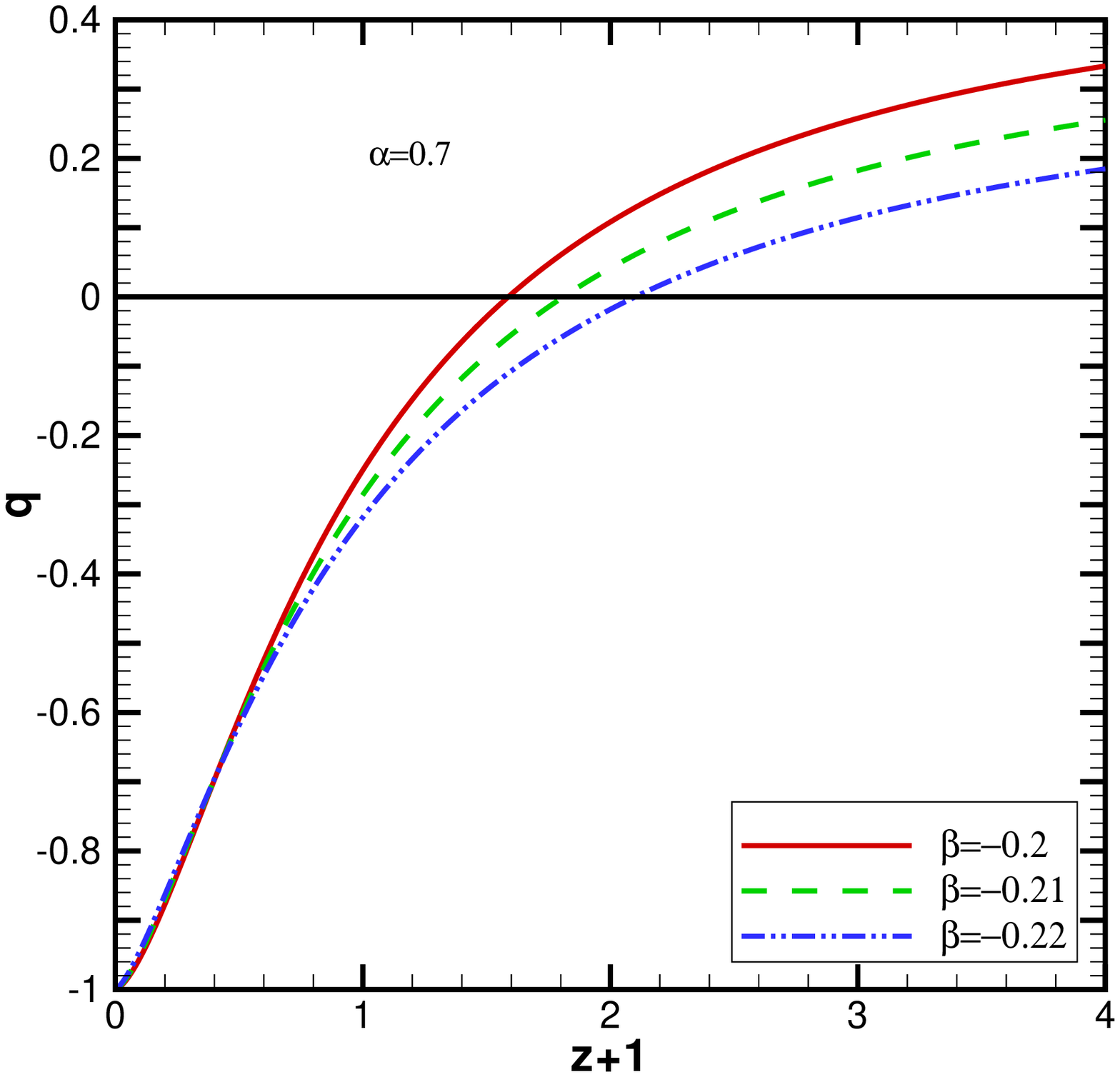}
\includegraphics[width=8cm]{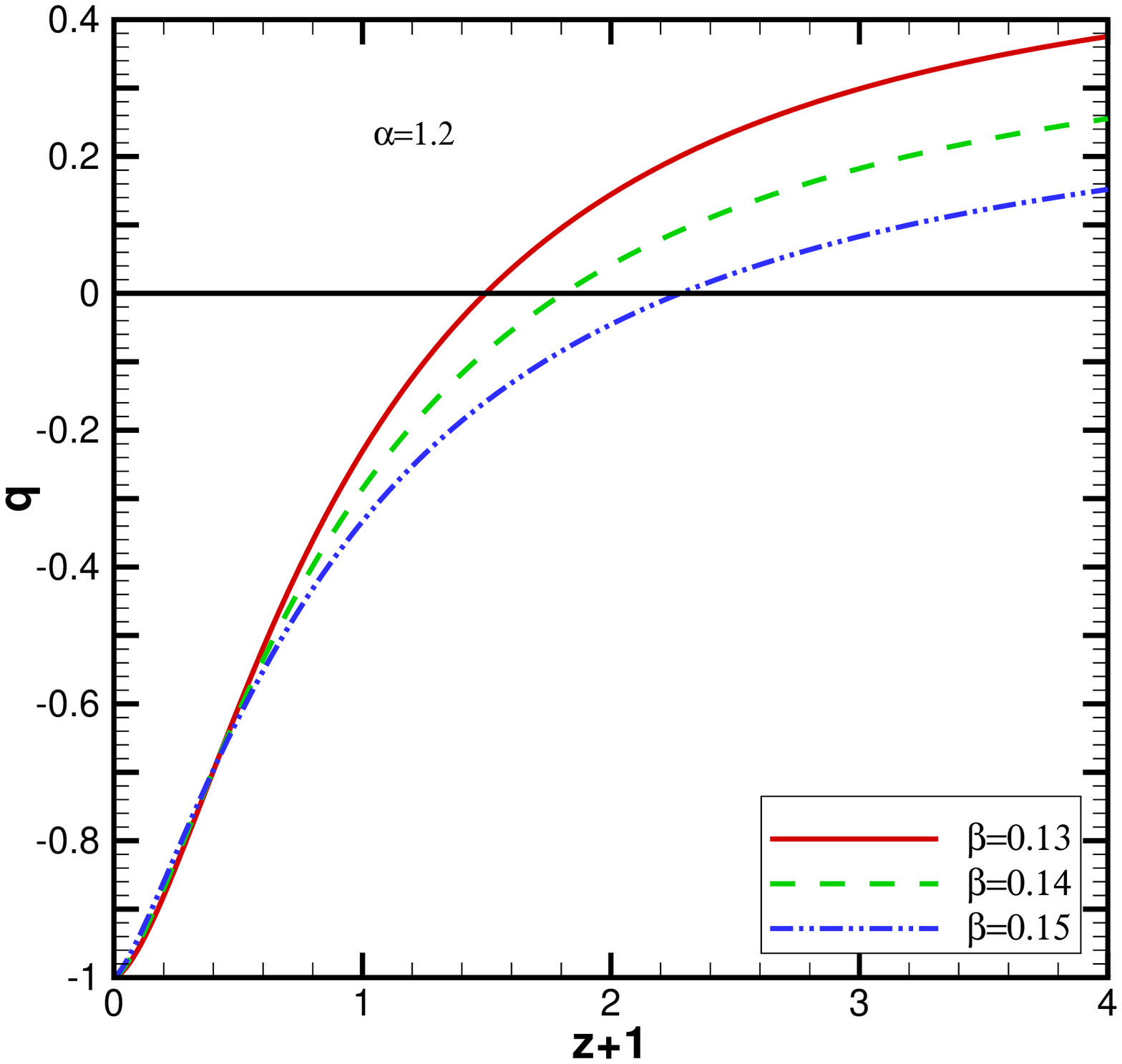}
\caption{The evolution of the deceleration parameter $q$ versus
redshift parameter $z$ for HDE with GO cutoff in DGP braneworld.
Left panel corresponds to $\alpha<1$ and $\beta<0$, while the
right panel shows the case $\alpha>1$ and $\beta>0$. }\label{q3}
\end{center}
\end{figure}
To check the limit of (\ref{wDz}) in standard cosmology where
$r_{c}\gg1$, we note from Eq. (\ref{scale}) and relation
$a/a_0=(1+z)^{-1}$ that $(a/a_0)^{\frac{\alpha-1}{\beta}}=(1+z)^{\frac{1-\alpha}{\beta}}\rightarrow0$,
as $r_{c}\gg1$. Therefore, (\ref{wDz}) reduces to
\begin{equation}\label{wz2}
\omega_D=-1+\frac{2}{3}\frac{\alpha-1}{\beta},
\end{equation}
which is exactly the result obtained in \cite{Granda}. It is
important to note that the EoS parameter of HDE with GO cutoff in
DGP braneworld scenario varies with time. Thus, it seems that the
presence of the extra dimension brings rich physics. For
$\alpha=1$ we have $\omega_D=-1$, similar to the cosmological
constant. The case with $\alpha<1$ and $\alpha>1$ should be
considered separately. In the first case where $\alpha<1$, we find
that the EoS parameter can explain the acceleration of the
Universe provided $\beta<0$. In the second case where $\alpha>1$,
the accelerated expansion can be achieved for $\beta>0$. In both
cases, our Universe has a transition from deceleration to the
acceleration phase around $0.4\leq z\leq1$, compatible with the
recent observations \cite{Daly,Kom1,Kom2}, and mimics the
cosmological constant at the late time. In figures (\ref{w3}) and
(\ref{q3}) we have shown the evolution of the EoS and the
deceleration parameter for HDE with GO cutoff in DGP braneworld.

\begin{figure}[htp]
\begin{center}
\includegraphics[width=8cm]{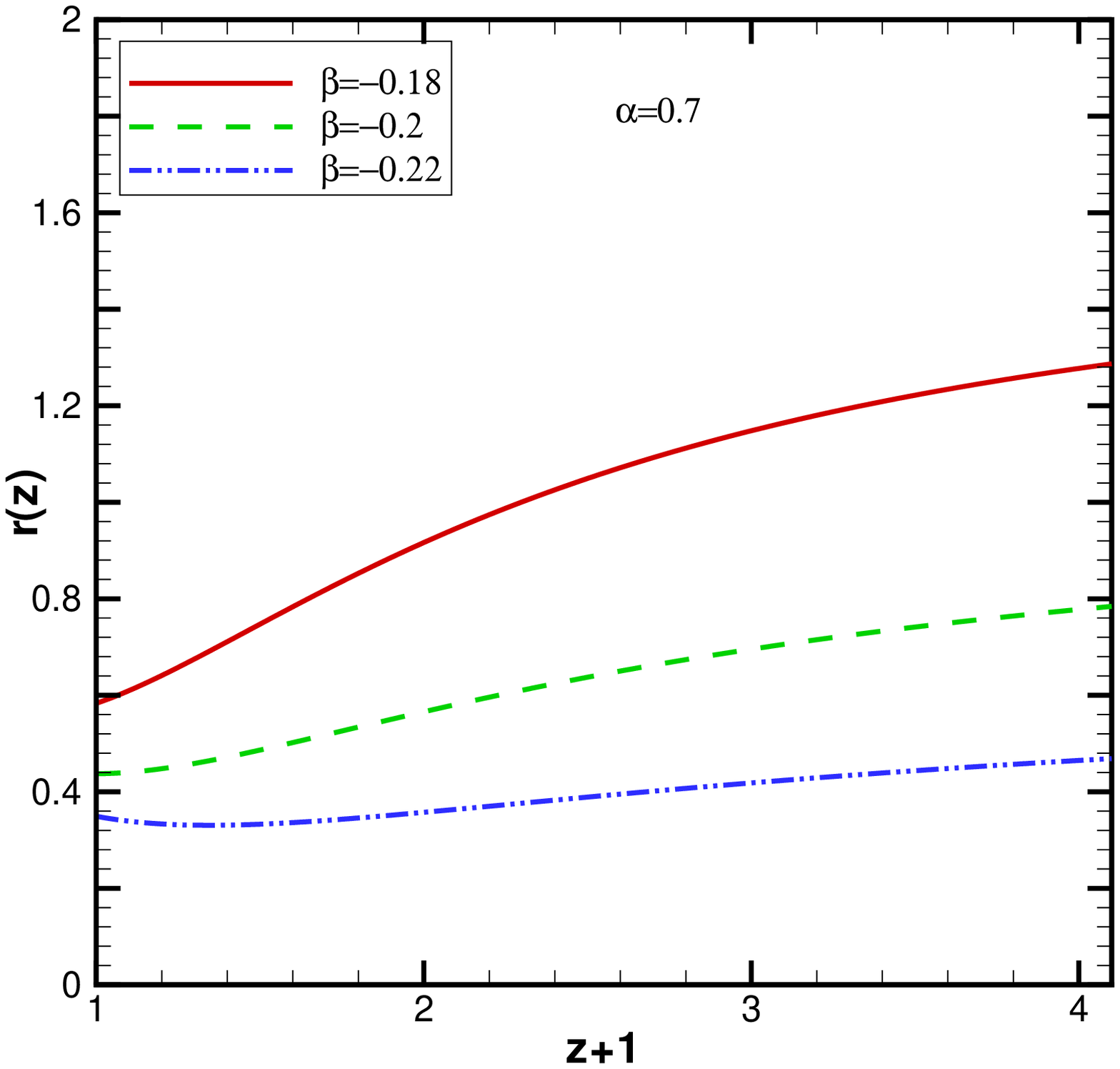}
\includegraphics[width=8cm]{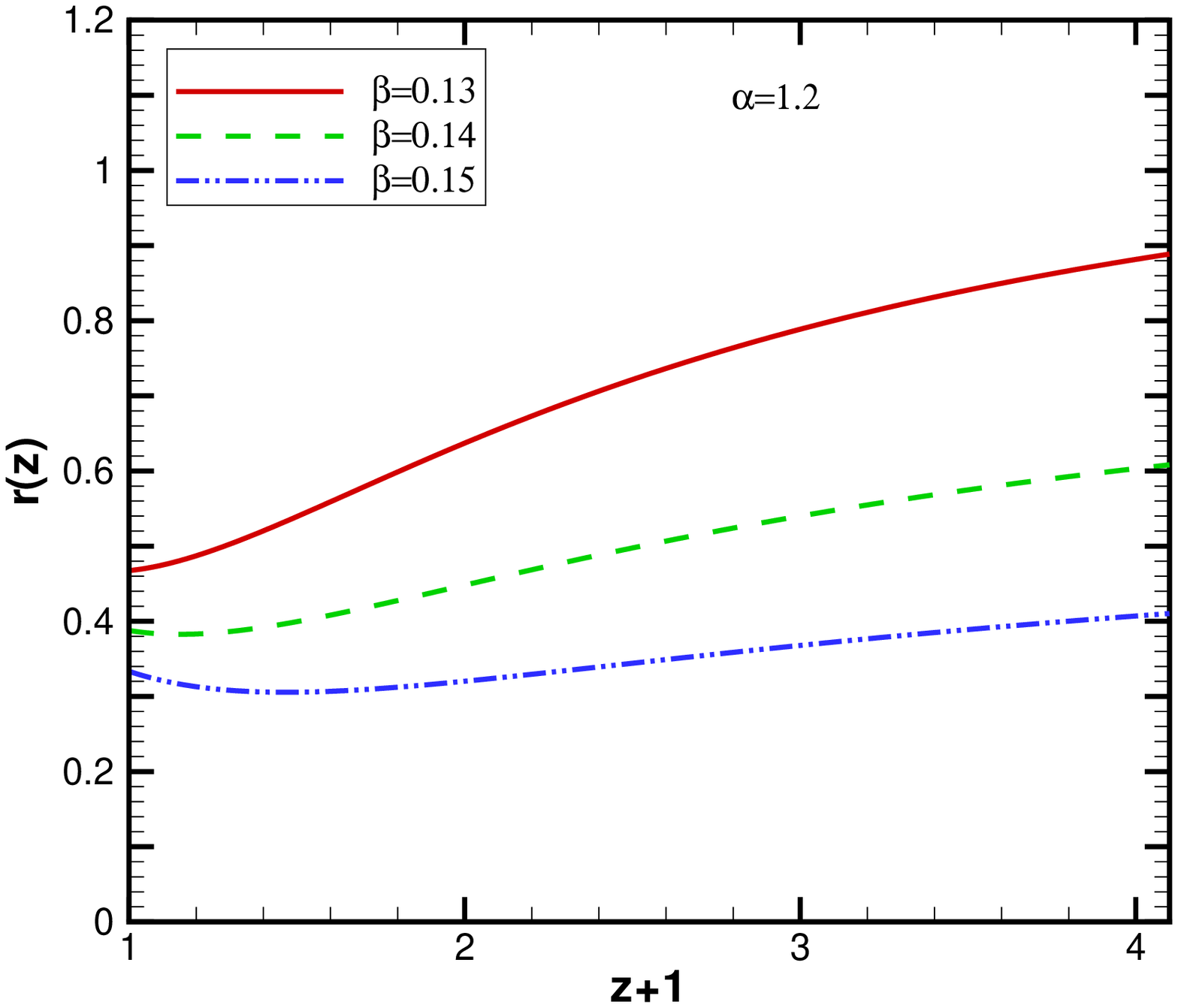}
\caption{The evolution of the statefinder parameter $r$ versus
$1+z$ for HDE with GO cutoff in DGP braneworld.}\label{r2}
\end{center}
\end{figure}

\begin{figure}[htp]
\begin{center}
\includegraphics[width=8cm]{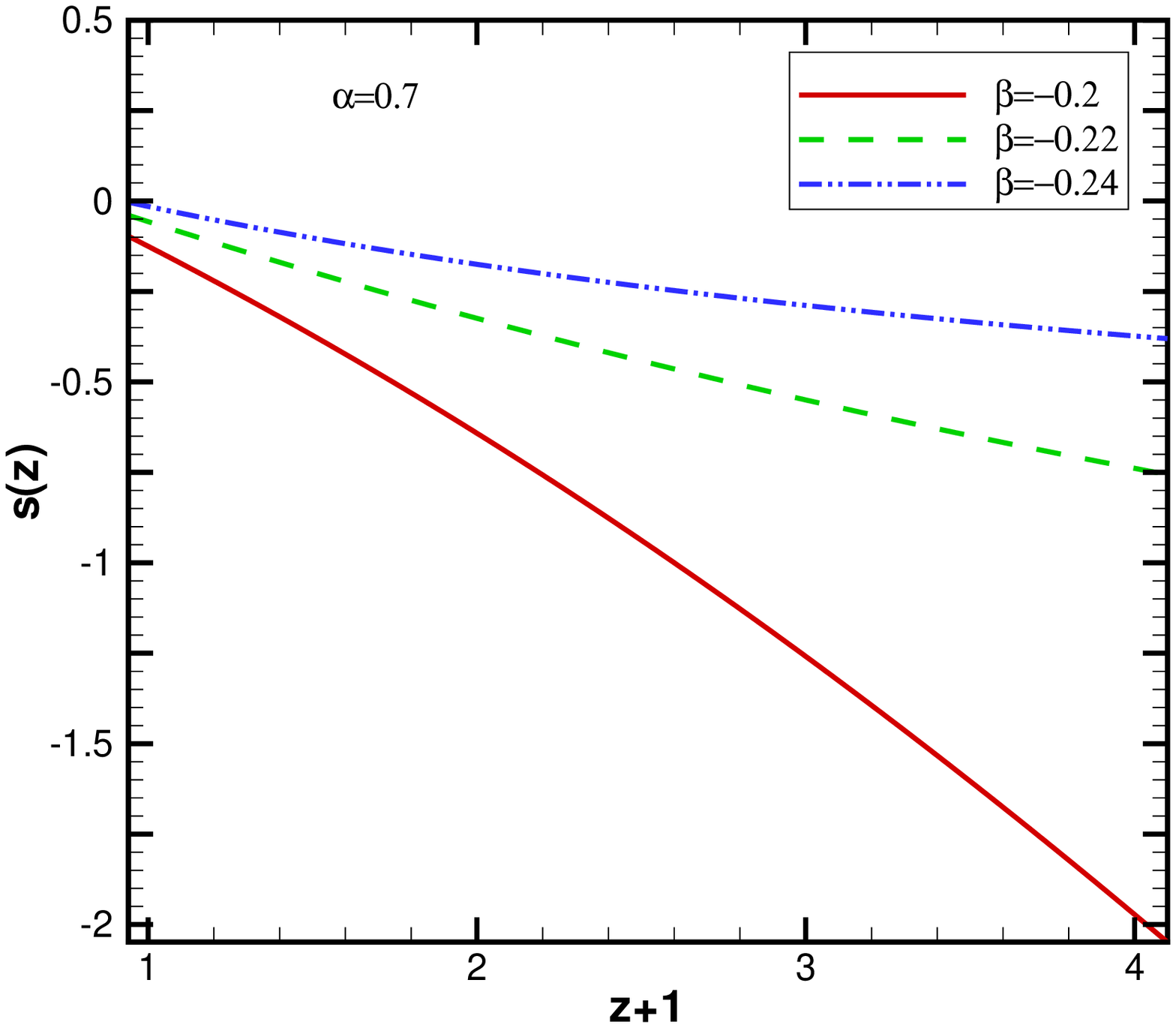}
\includegraphics[width=8cm]{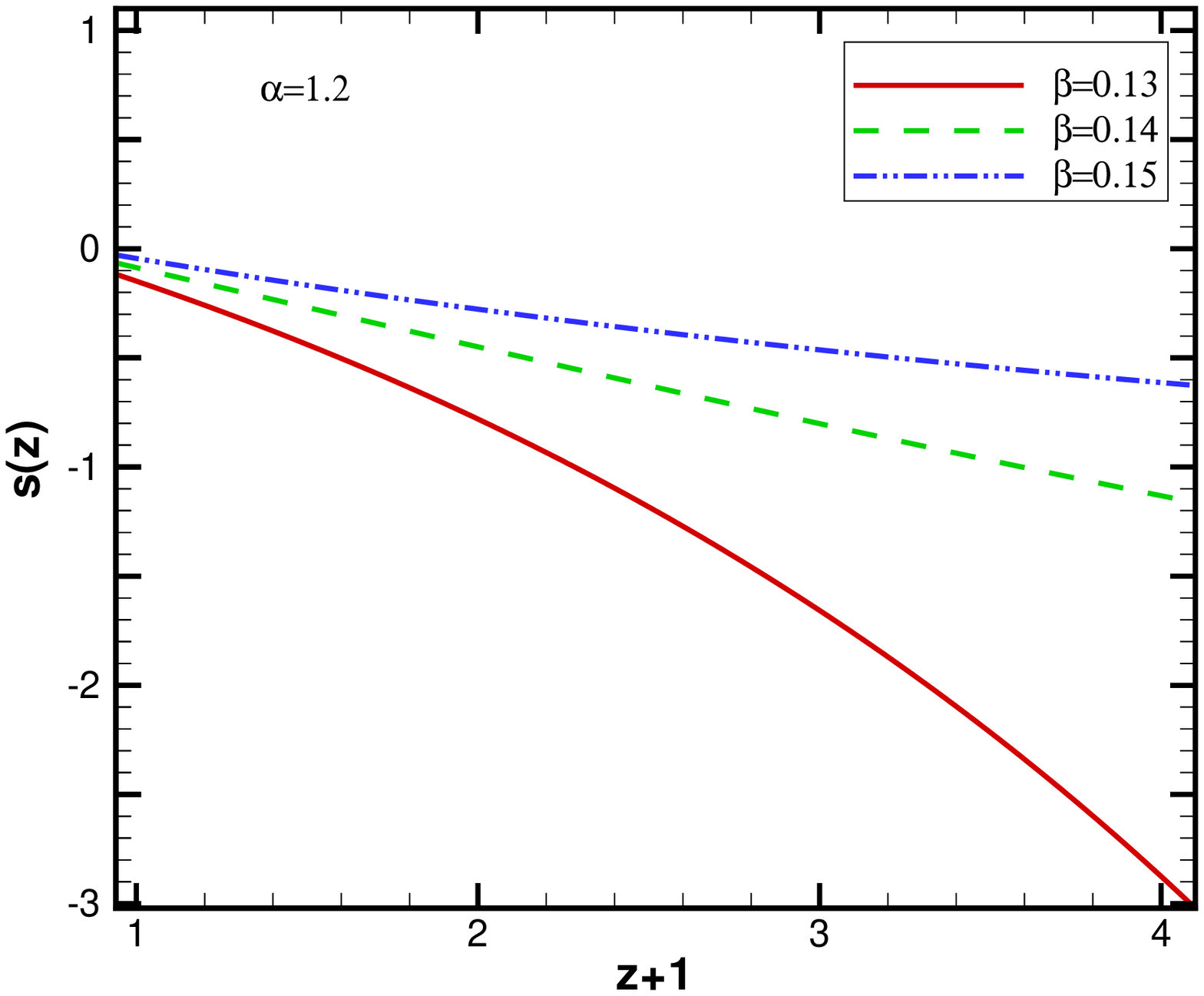}
\caption{The evolution of the statefinder parameter $s$ versus
$1+z$ for HDE with GO cutoff in DGP braneworld. }\label{s}
\end{center}
\end{figure}

\begin{figure}[htp]
\begin{center}
\includegraphics[width=8cm]{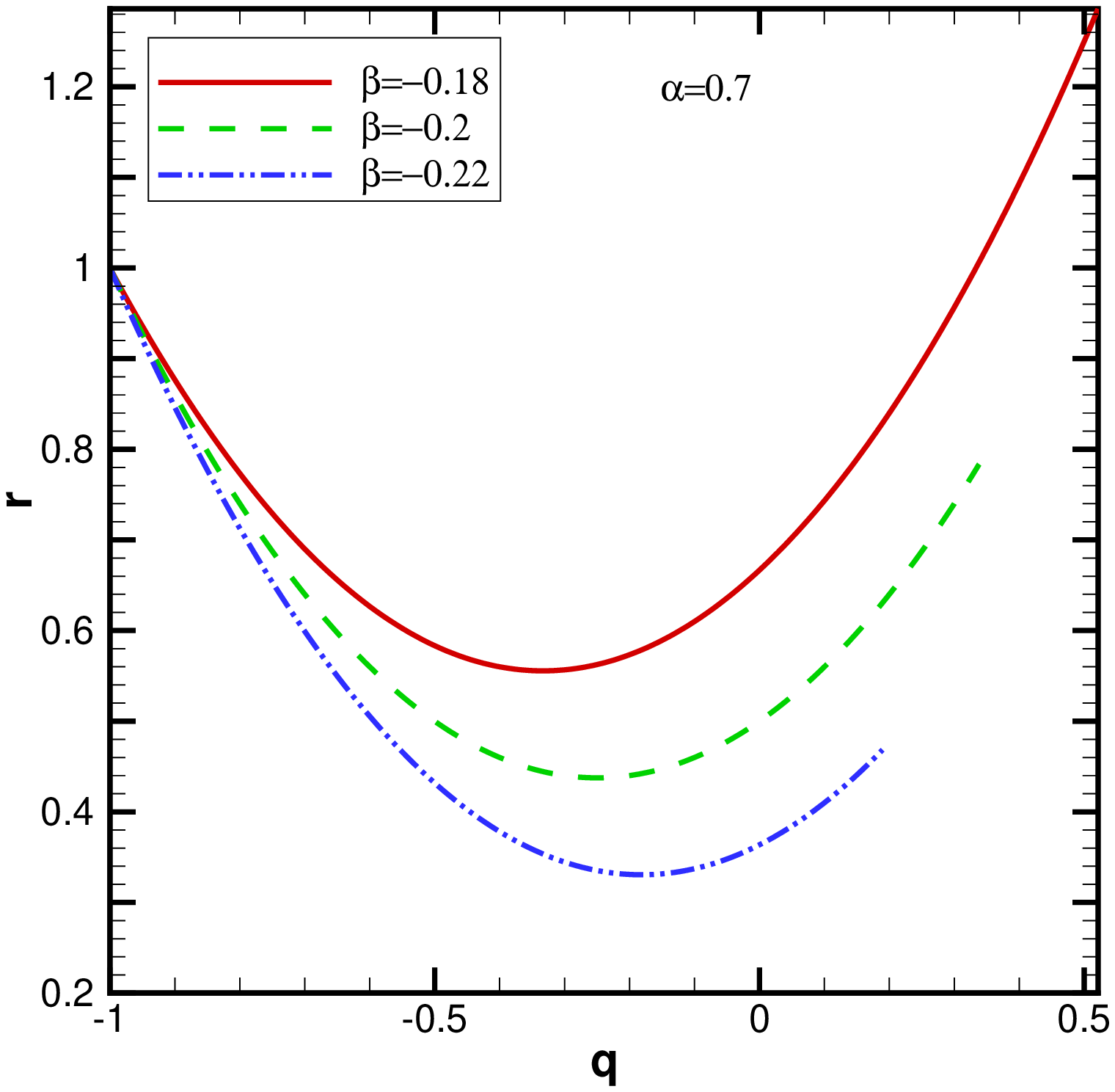}
\includegraphics[width=8cm]{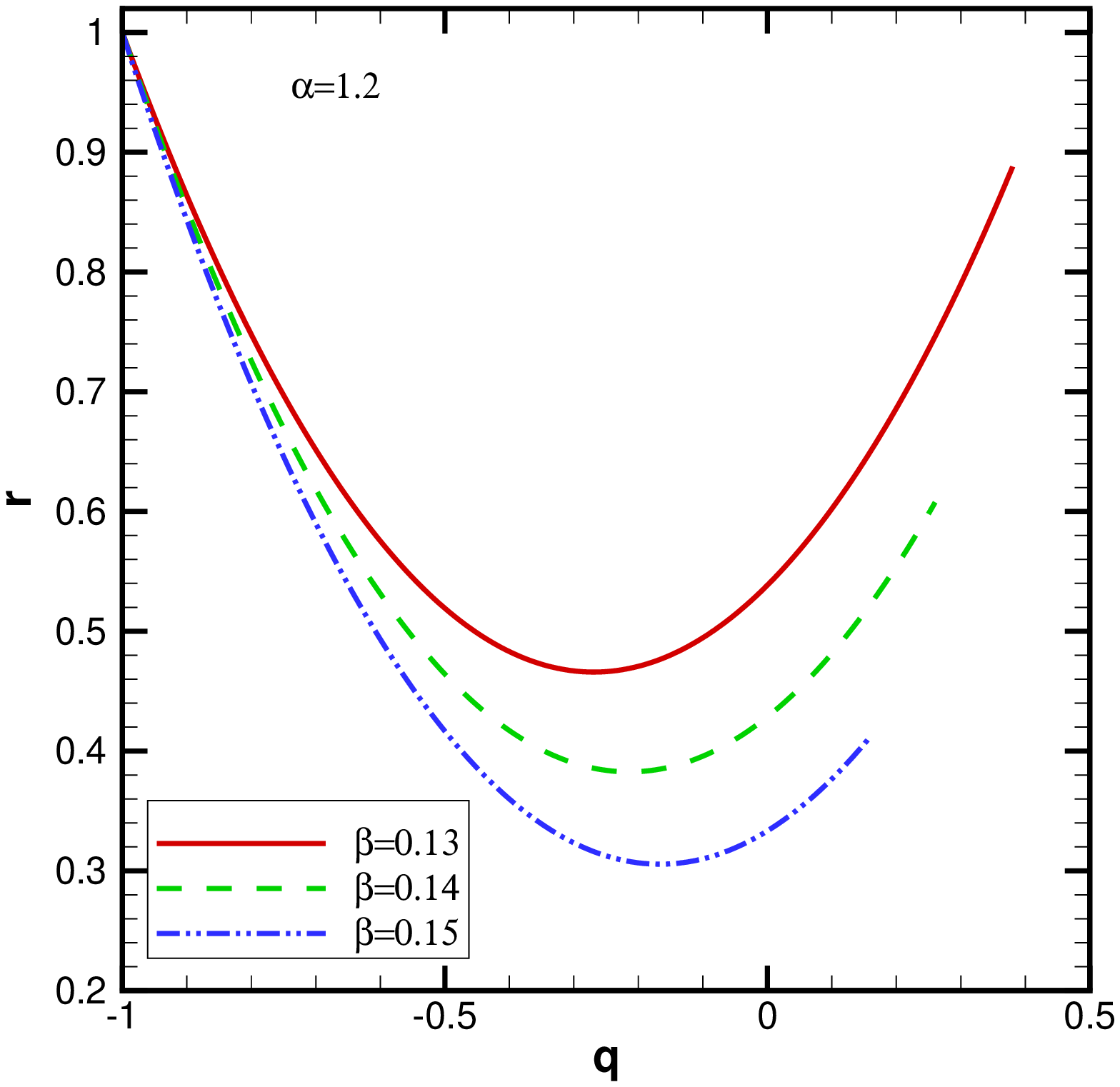}
\caption{The evolution of the statefinder parameter $r$ versus
deceleration parameter $q$ for HDE with GO cutoff in DGP
braneworld. }\label{r-q,DGP}
\end{center}
\end{figure}

\begin{figure}[htp]
\begin{center}
\includegraphics[width=8cm]{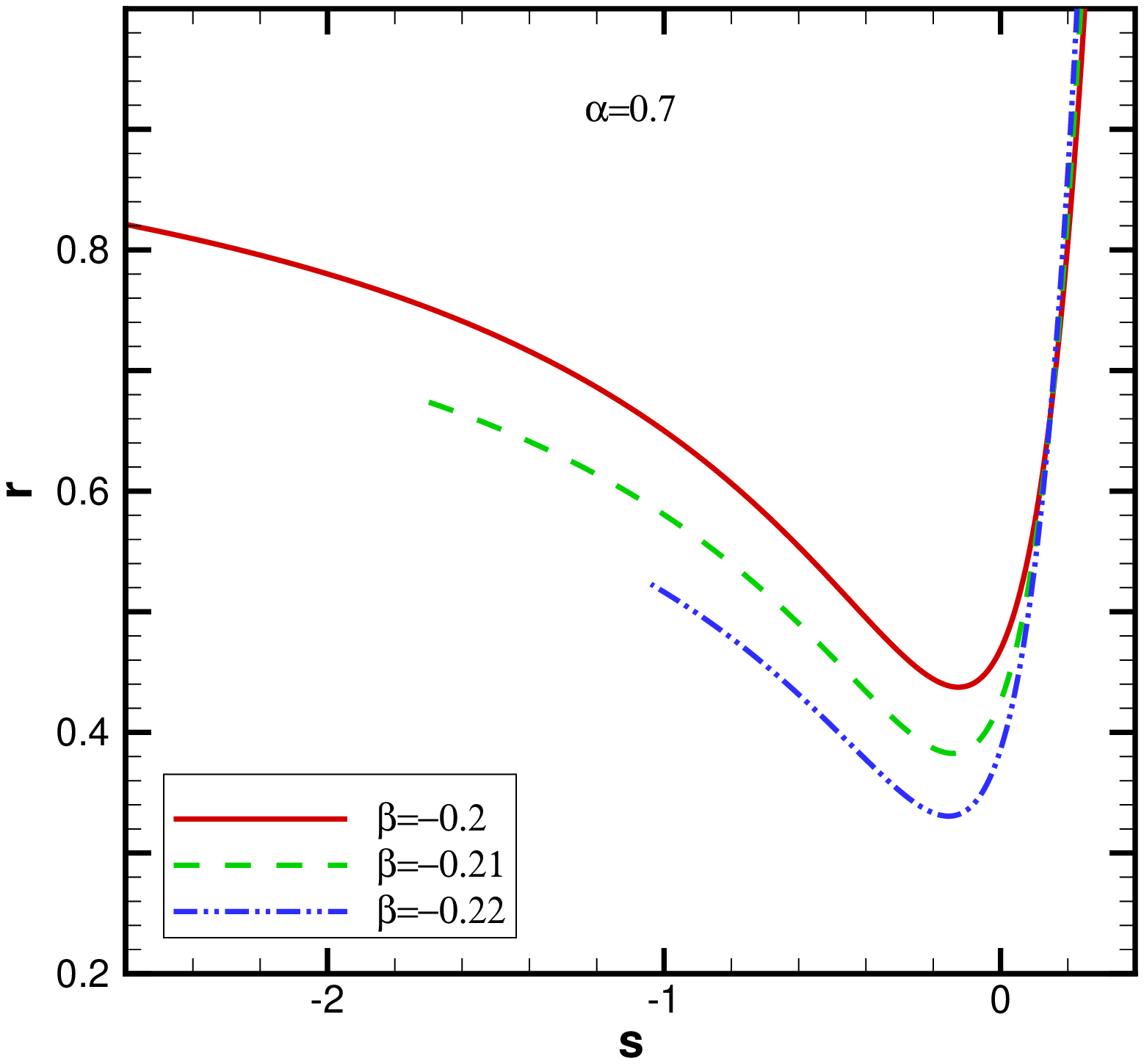}
\includegraphics[width=8cm]{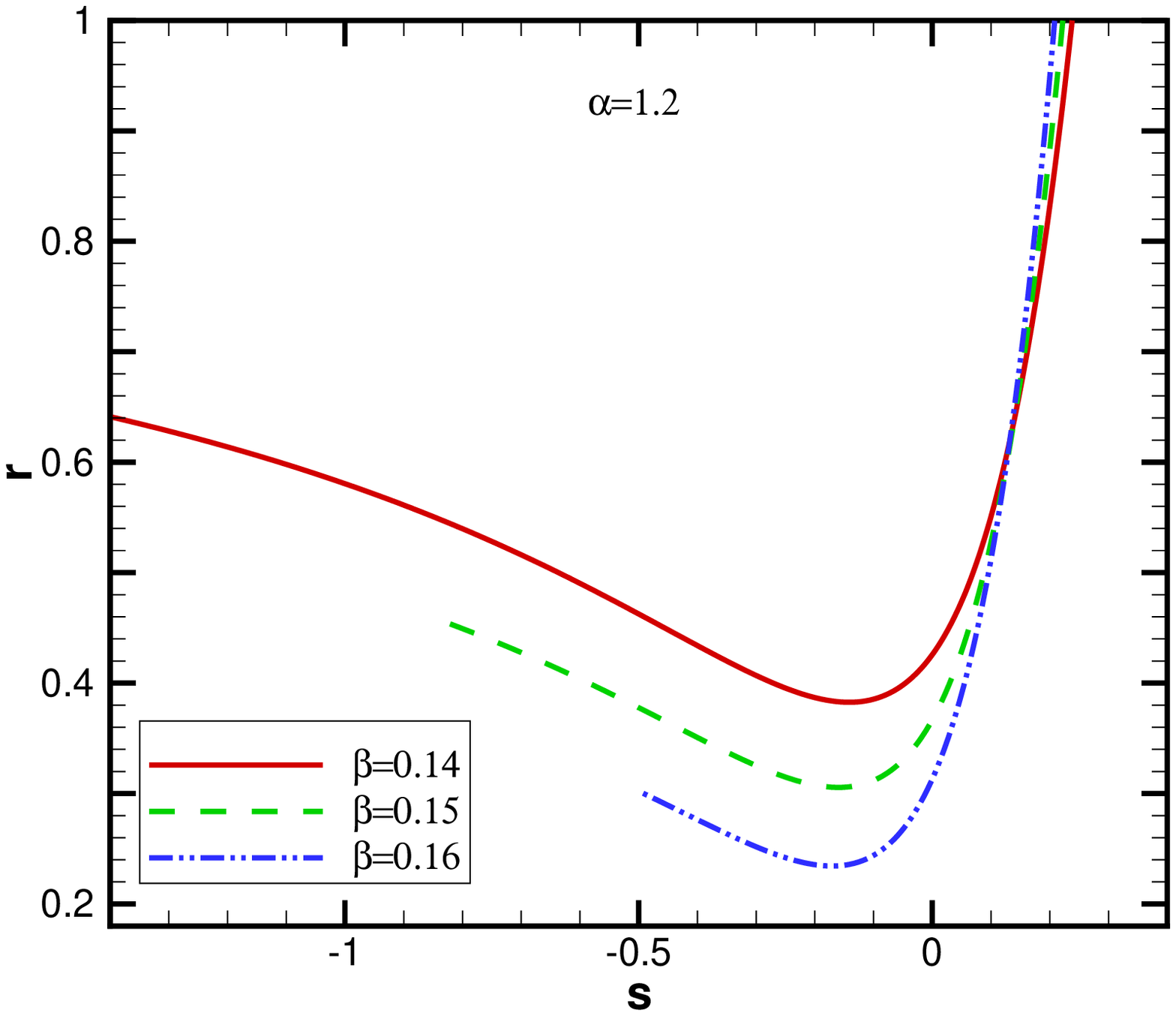}
\caption{The evolution of the statefinder parameter $r$ versus $s$
for for HDE with GO cutoff in DGP braneworld.}\label{r-s}
\end{center}
\end{figure}

Combining  Eqs. (\ref{H4}) and (\ref{H5}) with (\ref{r}) and
(\ref{rr2}), after using Eqs. (\ref{scale}) and (\ref{q1}), we
obtain the statefinder pair parameters as
\begin{eqnarray}\label{r4}
r&=&1+(1-\alpha)\Big[\frac{3\beta(a^{\frac{\alpha-1}{\beta}}+1)+(1-\alpha)(2+a^{\frac{\alpha-1}{\beta}})}{\beta^2(a^{\frac{\alpha-1}{\beta}}+1)^2}\Big],\\
s&=&\frac{\alpha-1}{6\beta}-\frac{(\alpha-1)^2}{3\beta}\frac{2({1+a^{\frac{\alpha-1}{\beta}}})+1}
{\Big(3\beta(1+a^{\frac{\alpha-1}{\beta}})+2(1-\alpha)\Big)(1+a^{\frac{\alpha-1}{\beta}})}.\label{r5}
\end{eqnarray}
For $\alpha=1$ we have $\{r,s\}=\{1,0\}$. In term of the redshift
parameter, $1+z=a^{-1}$, with $a_0=1$ for present time, Eqs.
(\ref{r4}) and (\ref{r5}) takes the form
\begin{eqnarray}
r&=&1+(1-\alpha)\Big[\frac{3\beta((1+z)^{\frac{1-\alpha}{\beta}}+1)+(1-\alpha)(2+(1+z)^{\frac{1-\alpha}{\beta}})}{\beta^2((1+z)^{\frac{1-\alpha}{\beta}}+1)^2}\Big],\\
s&=&\frac{\alpha-1}{6\beta}-\frac{(\alpha-1)^2}{3\beta}\frac{2({1+(1+z)^{\frac{1-\alpha}{\beta}}})+1}
{\Big(3\beta(1+(1+z)^{\frac{1-\alpha}{\beta}})+2(1-\alpha)\Big)(1+(1+z)^{\frac{1-\alpha}{\beta}})}.
\end{eqnarray}
The evolution of the statefinder pair parameters for HDE with GO
cutoff in the framework of DGP braneworld have been shown in
figures \ref{r2}-\ref{r-s}. Interestingly enough, from figure
(\ref{r-q,DGP}) we see that $r$ diverge as $q\rightarrow1/2$,
which corresponds to the matter dominated Universe, and mimics the
cosmological constant, namely $r=1, q=-1$, in the far future where
$z\rightarrow0$.

\section{Conclusions and discussion}
Following the discovery of the accelerated expansion of the
Universe, many DE models have been introduced to explain such a
cosmic acceleration. In order to discriminate various DE models,
the statefinder  diagnosis pair parameters $\{r, s\}$ was
introduced \cite{Sahni,Alam}. This diagnostic is constructed from
scale factor $a(t)$ and its derivatives up to the third order. It
was argued that trajectories in the $r-s$ plane correspond to a
wide variety of DE models. The spatially flat $\Lambda$CDM
scenario corresponds to a fixed point $\{r, s\} = \{1, 0\}$ in the
statefinder  diagnosis diagram. Departure of a given DE model from
this fixed point provides a good way for establishing the
statefinder of the presented model from $\Lambda$CDM.

In this paper, we have considered HDE in standard cosmology as
well as DGP braneworld. We have chosen the
Hubble radius $L=H^{-1}$ and the GO cutoff $L=(\alpha
\dot{H}+\beta H^2)^{-1/2}$, inspired by Ricci scalar curvature of
flat FRW universe, as the system's IR cutoff . We have investigated the evolution of the EoS,
deceleration and statefinder parameters. We found that in standard
cosmology for both proposed cutoff, the statefinder pair
parameters $\{r, s\}$ become constant depending on the model
parameters and with suitably choosing the parameters they restore
those of $\Lambda$CDM model. However, in the framework of DGP
braneworld, the situation become completely different and some
interesting results can be observed. Interestingly enough, we
found that although choosing $L=H^{-1}$ as the IR cutoff, in the
absence of interaction, leads to the wrong EoS of dust
$w_D=0$ \cite{Li}, here because of the effects of extra dimensions
the natural choice for IR cutoff in flat Universe, namely the
Hubble radius $L=H^{-1}$, can lead to an accelerated expanding Universe. In
this case $w_D$ and $q$, as well as the pair parameters $\{r, s\}$
depend on $\Omega_{r_c}={1}/({4H^2r_c^2})$. For GO cutoff in DGP
braneworld, the cosmological parameters can be obtained as
the function of redshift parameter $z$, and in case of $\alpha=0$, we
reach to $\{r, s\} = \{1, 0\}$, as expected. We have plotted the
behaviour of $\{r, s\}$ and observed that $r$ diverges at the past
as $ q\rightarrow1/2$, which corresponds to the matter dominated
era, and mimics the cosmological constant with $r=1, q=-1$ in the
future.

%%%%%%%%%%%%%%%%%%%%%%%%%%%%%%%%%%%%%%%%%%%%%%%%%%%%%%%%%%%%%%%%%%%%%%%%%%%%%%%%%%%%%%%%%%%%%%%%%
\acknowledgments{We thank referee for constructive comments. We
also thank Shiraz University Research Council. This work has been
supported financially by Research Institute for Astronomy \&
Astrophysics of Maragha (RIAAM), Iran.
}%%%%%%%%%%%%%%%%%%%%%%%%%%%%%%%%%%%%%%%%%%%%%%%


\begin{thebibliography}{99}
\bibitem{Riess} A.G. Riess, et al., Astron. J. {\bf 116} (1998)
1009;\\
S. Perlmutter, et al., Astrophys. J. {\bf 517} (1999) 565;\\
P. deBernardis, et al., Nature {\bf 404} (2000) 955;\\
S. Perlmutter,et al., Astrophys. J. {\bf 598} (2003) 102.


\bibitem{COL2001} M. Colless et al., Mon. Not. R. Astron. Soc. \textbf{328},
1039 (2001);\\ M. Tegmark et al., Phys. Rev. D \textbf{69}, 103501
(2004);\\ S. Cole et al., Mon. Not. R. Astron. Soc. \textbf{362},
505 (2005);\\ V. Springel, C.S. Frenk, and S.M.D. White, Nature
(London) \textbf{440}, 1137 (2006).

\bibitem{HAN2000} S. Hanany et al., Astrophys. J. Lett. \textbf{545}, L5
(2000);\\ C.B. Netterfield et al., Astrophys. J. \textbf{571}, 604
(2002);\\ D.N. Spergel et al., Astrophys. J. Suppl. \textbf{148},
175 (2003).

\bibitem{Wetterich} C. Wetterich, Nucl. Phys. B {\bf 302} (1988) 668;\\
B. Ratra, J. Peebles, Phys. Rev. D {\bf 37} (1988) 321.

\bibitem{Caldwell} R.R. Caldwell, Phys. Lett. B {\bf 545} (2002) 23;\\
S. Nojiri, S.D. Odintsov, Phys. Lett. B {\bf 562} (2003) 147;\\
S. Nojiri, S.D. Odintsov, Phys. Lett. B {\bf 565} (2003) 1.
\bibitem{Chiba} T. Chiba, T. Okabe, M. Yamaguchi, Phys. Rev. D {\bf 62} (2000) 023511;\\
C. Armendáriz-Picón, V. Mukhanov, P.J. Steinhardt, Phys. Rev. Lett. {\bf 85} (2000)\\
4438;\\
C. Armendáriz-Picón, V. Mukhanov, P.J. Steinhardt, Phys. Rev. D {\bf 63} (2001)
103510.

\bibitem{Kamenshchik} A. Kamenshchik, U. Moschella, V. Pasquier, Phys. Lett. B 511 (2001) 265;\\
M.C. Bento, O. Bertolami, A.A. Sen, Phys. Rev. D 66 (2002) 043507.

\bibitem{Cai1} R.G. Cai, Phys. Lett. B 657 (2007) 228;\\
H. Wei, R.G. Cai, Phys. Lett. B {\bf 660} (2008) 113;\\
K.Y. Kim, H.W. Lee, Y.S. Myung, Phys. Lett. B {\bf 660} (2008) 118;\\
H. Wei, R.G. Cai, Phys. Lett. B {\bf 663} (2008) 1;\\
J.P. Wu, D.Z. Ma, Y. Ling, Phys. Lett. B {\bf 663} (2008) 152;\\
J. Zhang, X. Zhang, H. Liu, Eur. Phys. J. C {\bf 54} (2008) 303;\\
H. Wei, R.G. Cai, Eur. Phys. J. C {\bf 59} (2009) 99;\\
I.P. Neupane, Phys. Lett. B {\bf 673} (2009) 111.

\bibitem{Shey2} A. Sheykhi, Phys. Lett. B {\bf680}, 113 (2009)%Interacting agegraphic dark energy models in non-flat Universe%
;\\ A. Sheykhi, Phys. Lett. B {\bf682}, 329 (2010) 329 %ADE tachyon %
;\\ A. Sheykhi, Phys. Rev. D {\bf81}, 023525 (2010);\\ Ahmad
Sheykhi, Mubasher Jamil, Phys. Lett. B {\bf694}, 284 (2011);\\
%Interacting HDE and NADE in Brans-Dicke Chameleon Cosmology Cited by 22 records%
A. Sheykhi, M. R. Setare, Int. J. Theor. Phys. {\bf49}, 2777 (2010). %Interacting new agegraphic viscous dark energy with varying%


\bibitem{Cohen1} A. Cohen, D. Kaplan, A. Nelson, Phys. Rev. Lett. {\bf 82} (1999)
4971.


\bibitem{Hsu} S. D. H. Hsu, Phys. Lett. B {\bf 594}, 13 (2004)
\bibitem{Li} M. Li, Phys. Lett. B {\bf 603}, 1 (2004).
\bibitem{pav1} D. Pavon, W. Zimdahl, Phys. Lett. B \textbf{628}, 206 (2005);\\
A. Sheykhi, Phys. Rev.  D {\bf84}, 107302 (2011).
\bibitem{Shey1} A. Sheykhi, Class. Quantum Grav. {\bf27},
025007 (2010).

\bibitem{Horava} P. Horava and D. Minic, Phys. Rev. Lett. {\bf 85}, 1610 (2000);
\\S. D. Thomas, Phys. Rev. Lett. {\bf 89}, 081301 (2002).
.

\bibitem{Fischler} W. Fischler and L. Susskind, hep-th/9806039;\\ R. Bousso, JHEP {\bf 9907},
004 (1999).
\bibitem{Nojiri} S. Nojiri and S. D. Odintsov, Gen. Rel. Grav. {\bf 38}, 1285 (2006).


\bibitem{Gao}C. J. Gao, X. L. Chen and Y. G. Shen, Phys. Rev. D {\bf 79}, 043511 (2009);\\
R. G. Cai, B. Hu and Y. Zhang, Commun. Theor. Phys. {\bf 51}, 954
(2009)
\bibitem{Granda} L.N. Granda, A. Oliveros, Phys. Lett. B {\bf 669} (2008)
275;\\ L.N. Granda, A. Oliveros, Phys. Lett. B {\bf671} (2009)
199.

\bibitem{Sahni} V. Sahni, T.D. Saini, A. A. Starobinski and U. Alam, JETP Lett. {\bf 77}, 201 (2003).
\bibitem{Alam} U. Alam, V. Sahni, T.D. Saini and A. A. Starobinski, Mon. Not. R. Astron. Soc. {\bf 344}, 1057 (2003).

\bibitem{Feng} C.J. Feng, Phys. Lett. B {\bf670}, 231 (2008).

\bibitem{state} J. Zhang, X.
Zhang, H. Liu, Phys. Lett. B 659 (2008) 26;\\ M. Malekjani, A.
Khodam-Mohammadi, N. Nazari-pooya, Astrophys.Space Sci. 332 (2011)
515;\\ M.R. Setare, M. Jamil, Gen. Relativ. Gravit. 43 (2011)
293;\\ L. Zhang, J. Cui, J. Zhang, X. Zhang, Int. J. Mod. Phys.D
19 (2010) 21.

\bibitem{state1} Fei Yu, Jing-Fei Zhang, Commun. Theor. Phys. 59 (2013) 243;\\
Jing-Lei Cui, Jing-Fei Zhang, Eur. Phys. J. C 74 (2014)
2849. %Comparing holographic dark energy models with statefinder%
\bibitem{Bertolami} O. Bertolami, F. Gil Pedro, and M. Le Delliou. Phys. Lett. B,
{\bf 654};\\
 M. Baldi. Mon. Not. R. Astron. Soc. {\bf 414}, 116 (2011).
\bibitem{Hu}  B. Hu, Y. Ling,  Phys. Rev. D {\bf 73}, 123510 (2006).

\bibitem{Amendola}L. Amendola, Phys. Rev. D {\bf 60} 043501 (1999) ;\\
L. Amendola, Phys. Rev. D {\bf 62} (2000) 043511;\\
L. Amendola and C. Quercellini, Phys. Rev. D {\bf 68} 023514 (2003) ;\\
L. Amendola and D. Tocchini-Valentini, Phys. Rev. D {\bf 64} 043509 (2001)  ;\\
L. Amendola and D. T. Valentini, Phys. Rev. D {\bf 66} 043528
(2002) .

\bibitem{Zimdahl}W. Zimdahl and D. Pavon, Phys. Lett. B {\bf 521} 133 (2001);\\
W. Zimdahl and D. Pavon, Gen. Rel. Grav. {\bf 35} 413 (2003).


\bibitem {GO} M. Jamil, K. karami, A. Sheykhi, E. Kazemi, Z. Azarmi, Int.
J. Theor. Phys. {\bf51}, 604 (2012);\\ S. Ghaffari, M. H. Dehghani
and A. Sheykhi, Submitted to Phys. Rev. D (2014).

\bibitem{Dvali} G.R. Dvali, G. Gabadadze, and M. Porrati, Phys.Lett. B {\bf 485} 208 (2000);\\
Deffayet,D., Phys.Lett. B {\bf 502} 199 (2001);\\
Deffayet,D., Dvali, G.R. and Gabadadze,G., Phys.Rev. D {\bf 65}
044023 (2002).
\bibitem{Dvali2} G. Dvali, G. Gabadadze, Phys. Rev. D {\bf 63}, 065007 (2001).
\bibitem{Alder} S.L. Adler, Phys. Rev. Lett. {\bf 44} (1980) 1567;\\
S.L. Adler, Phys. Lett. B {\bf 95} (1980) 241;\\
S.L. Adler, Rev. Mod. Phys. {\bf 54} (1982) 729;\\
D.M. Capper, Nuovo Cimento A {\bf 25} (1975) 29;\\
A. Zee, Phys. Rev. Lett. {\bf 48} (1982) 295.

\bibitem{Def}  C. Deffayet, Phys. Lett. B \textbf{502}, 199 (2001).
\bibitem{Gaffari} S. Ghaffari, M. H. Dehghani, A. Sheykhi, Phys.
Rev. D {\bf 89}, 123009 (2014).
\bibitem{Koyama} K. Koyama, Class. Quant. Grav. {\bf 24}, R 231 (2007).
\bibitem{Cai2} Y.F. Cai et al, Physics Reports {\bf 493}, 1 (2010).

\bibitem{Deffayet1} C. Deffayet, G.R. Dvali, G. Gabadadze, Phys. Rev. D 65 (2002) 044023;\\
V. Sahni, Y. Shtanov, J. Cosmol, Astropart. Phys. 11 (2003) 014.

\bibitem{Daly} R.A. Daly et al., Astrophys. J. {\bf677}, 1 (2008).
\bibitem{Kom1} E. Komatsu et al. [WMAP Collaboration], Astrophys. J. Suppl. \textbf{192}, 18
(2011).

 \bibitem{Kom2} V. Salvatelli, A. Marchini, L. L. Honorez and O. Mena, Phys.
Rev. D {\bf88}, 023531 (2013).

\end{thebibliography}
\end{document}